\documentclass[preprint,superscriptaddress,preprintnumbers,amsmath,amssymb,prb]{revtex4}
\usepackage{graphicx}
\usepackage{epstopdf}
\usepackage{threeparttable}
\usepackage[version=3]{mhchem}

\usepackage{subfigure}
\usepackage{bm}

\begin{document}

\title{All-electron self-consistent GW in the Matsubara-time domain: implementation and benchmarks of semiconductors and insulators}

\author{Iek-Heng Chu} 
\affiliation{Department of Physics and Quantum Theory Project, University of Florida, Gainesville, Florida 32611, United States}
\author{Jonathan P. Trinastic} 
\affiliation{Department of Physics and Quantum Theory Project, University of Florida, Gainesville, Florida 32611, United States}
\author{Yun-Peng Wang} 
\affiliation{Department of Physics and Quantum Theory Project, University of Florida, Gainesville, Florida 32611, United States}
\author{Adolfo G. Eguiluz}
\affiliation{Department of Physics and Astronomy, The University of Tennessee, Knoxville, Tennessee 37996, USA}
\author{Anton Kozhevnikov}
\affiliation{Institute for Theoretical Physics, ETH Zurich, 8093 Zurich, Switzerland}
\author{Thomas C. Schulthess}
\affiliation{Institute for Theoretical Physics, ETH Zurich, 8093 Zurich, Switzerland}
\author{Hai-Ping Cheng}
\email{cheng@qtp.ufl.edu}
\affiliation{Department of Physics and Quantum Theory Project, University of Florida, Gainesville, Florida 32611, United States}

\begin{abstract}
{The GW approximation is a well-known method to improve electronic structure predictions calculated within density functional theory. In this work, we have implemented a computationally efficient GW approach that calculates central properties within the Matsubara-time domain using the modified version of Elk, the full-potential linearized augmented plane wave (FP-LAPW) package. Continuous-pole expansion (CPE), a recently proposed analytic continuation method, has been incorporated and compared to the widely used Pade approximation. Full crystal symmetry has been employed for computational speedup. We have applied our approach to 18 well-studied semiconductors/insulators that cover a wide range of band gaps computed at the levels of single-shot G$_0$W$_0$, partially self-consistent GW$_0$, and fully self-consistent GW (scGW). Our calculations show that G$_0$W$_0$ leads to band gaps that agree well with experiment for the case of simple $s$-$p$ electron systems, whereas scGW is required for improving the band gaps in 3-$d$ electron systems. In addition, GW$_0$ almost always predicts larger band gap values compared to scGW, likely due to the substantial underestimation of screening effects. Both the CPE method and Pade approximation lead to similar band gaps for most systems except strontium titantate, suggesting further investigation into the latter approximation is necessary for strongly correlated systems. Our computed band gaps serve as important benchmarks for the accuracy of the Matsubara-time GW approach. }
\end{abstract}

\maketitle

\section{Introduction} 
Calculations using density functional theory\cite{Kohn1965,Hohenberg1964} (DFT) have become the standard \textit{ab initio} technique to study the electronic and structural properties of molecules, nanoparticles, and periodic solids.\cite{Kohn1996,Dreizler2012,Norskov2011,Ruiz2012}  However, it is well-known that the electronic band gap of semiconductors and insulators is severely underestimated within DFT due to the lack of a derivative discontinuity in standard exchange-correlation potentials.\cite{Sham1983}  This deficiency hinders the theory's useful application in fields such as optics, photovoltaics, thermoelectrics, and transport that require an accurate characterization of excited state properties.  

The GW approximation, originally proposed by Hedin,\cite{Hedin1965} provides a route to improve electronic descriptions and band gap results using many-body perturbation theory.  The central quantity in this approach is the exchange-correlation self-energy ($\Sigma^{xc}$), which incorporates (i) the exact electronic exchange interaction, and (ii) the complex electron-electron correlation accounting for screening effects often treated within the random phase approximation (RPA).\cite{Eshuis2012,Ren2012}  This approach has been applied to a wide variety of materials and provides electronic structure results in better agreement with experiments compared to its DFT counterpart.\cite{Chu2014,Yang2007,Shishkin2007,Kresse2012,Fuchs2007,Faleev2004}

Although studies employing the GW approximation have enjoyed early success in improving band gap predictions, many implementations rely on the pseudopotential (PP) approximation that treats pseudo wave functions and valence-core interactions at the level of DFT.\cite{Deslippe2012,Marini2001,Dixit2011,Pham2013} To avoid the PP approximation, several all-electron GW implementations have been reported in recent years based on the full-potential linearized augmented plane wave (FP-LAPW),\cite{Ku2002,Sharma2005,Friedrich2010,Jiang2013,Usuda2002} the linearized muffin-tin orbital (LMTO),\cite{Usuda2002} and the projector-augmented wave (PAW)\cite{Bochl1994} in conjunction with a plane-wave basis.\cite{Shishkin2006,Shishkin2007}  Most of these all-electron studies have only implemented the G$_0$W$_0$ approximation due to its lower computational cost,\cite{Usuda2002,Gomez2008,Li2012} however these single-shot calculations are plagued by violations of momentum, energy, and particle conservation laws.\cite{Baym1961,Baym1962,Dahlen2006}  They also introduce a troubling dependence on the choice of Kohn-Sham (K-S) basis used as a zeroth order starting point.\cite{Rinke2005,Fuchs2007}  Fully self-consistent GW (scGW) calculations avoid these issues and provide an unbiased physical picture predicted by GW theory.  To date, few studies have performed scGW calculations within an all-electron framework,\cite{Carus2012,Caruso2013,Ku2002} among them includes the self-consistent GW method performed within the Matsubara-time domain,\cite{Mahan2013,Bruus2004,Ku2002} as first implemented by Ku and Eguiluz.\cite{Ku2002} However, this approach has only been applied to bulk Si and Ge and its applicability to other semiconductors and insulators requires further examination. 

There are two main advantages of performing GW calculations within the Matsubara-time domain.  First, $\Sigma$ is simply the product of the single-particle Green's function ($G$) and screened Coulomb interaction ($W$). In contrast, the solution for $\Sigma$ in Matsubara-frequency space requires a convolution of $G$ and $W$ that usually demands more frequency points to reach convergence.\cite{Kutepov2009} Second, the Green's function in Matsubara-time lacks singular points that can arise in frequency space, which leads to smoother single-particle Green's functions compared to those in the frequency domain. Despite these advantages, the need for a reliable analytic continuation technique makes accurate calculations within Matsubara-time particularly challenging. The Pade approximation is often adopted for this purpose due to its simple implementation and low computational efficiency.\cite{Vidberg1977} In this approach, the quantities of interest (e.g., $\Sigma$ and $G$) are expressed as fractional polynomials that are fitted to computed values in the Matsubara-frequency domain. Such expressions are then analytically continued into the real-frequency domain. The reliability of this approximation remains under debate, and recently Staar and co-workers have proposed the continuous-pole expansion (CPE) as an alternative algorithm for analytic continuation from the Matsubara-frequency to the real-frequency domain.\cite{Staar2014} Unlike the Pade approximation, this method explicitly takes into account the physical causality that places a constraint on the self-energy.

In this paper, we build upon an all-electron GW code we have already developed\cite{Kozhevnikov2010,Chu2014} by calculating $\Sigma$ within the Matsubara-time domain, which improves the code's computational efficiency and provides scGW calculations. We implement this method in conjunction with the CPE to solve for the quasiparticle energies in the real-frequency domain. We validate this method by investigating the electronic band gaps of a wide range of semiconductors and insulators at different levels of GW approximation. Our calculations demonstrate that the band gaps for 3-$d$ electron systems are often in better agreement with experiment when using scGW than the commonly used G$_0$W$_0$ approximation, whereas the latter approximation often yields reasonable experimental agreement in simple $s$-$p$ electron systems. We also find that both the CPE and Pade approximation yield very similar electronic band gaps among most tested systems, however the CPE method provides a better electronic description of strongly-correlated strontium titanate. 

The rest of the paper is organized as follows.  Section~\ref{theory} outlines the scGW approximation and Section~\ref{dev} describes its implementation within the existing all-electron DFT package. Results and discussion are then presented in Section~\ref{results}, followed by the conclusion in Section~\ref{conclusion}.

\section{Basics of the Theory\label{theory}}
Within the single-particle picture, the excitation properties of solids can be determined by the single-particle Green's function via Dyson equation. When expressed in real-space and Matsubara-time domain, the Dyson equation reads
\begin{eqnarray}
G({\bf r}, {\bf r}'| \tau)&&=G^0({\bf r}, {\bf r}'| \tau)+\int^{\beta}_{0}d\tau_1\int^{\beta}_0d\tau_{2}\int d{\bf r}_1\int d{\bf r}_2 G^0({\bf r},{\bf r}_1|\tau-\tau_1)\label{dysonGF}\\\nonumber
&& \times\Delta\Sigma({\bf r}_1,{\bf r}_2|\tau_1-\tau_2) G({\bf r}_2, {\bf r}'| \tau_2),
\end{eqnarray}
where $G$ and $G^0$ are the Green's functions associated with the interacting system of interest and a pre-selected reference system, respectively. In this work, the non-interacting K-S system calculated within DFT is adopted as the reference system. $\tau$ is the Matsubara-time argument that in general falls within [-$\beta$, $\beta$] where $\beta=1/k_BT$, $k_B$ is Boltzmann's constant, and $T$ is the temperature. Given that $G$ obeys the relation $G({\bf r}, {\bf r}', -\tau)=-G({\bf r}, {\bf r}', -\tau+\beta)$ for $\tau\in[0, \beta]$, it is sufficient to restrict our study to $\tau\in[0,\,\beta$]. $\Delta\Sigma$ is the change in the electron-electron interaction between the interacting and reference K-S systems:
\begin{eqnarray}
&&\Delta\Sigma({\bf r},{\bf r}'| \tau)=\Sigma({\bf r}, {\bf r}'|\tau) - \Sigma^0({\bf r}, {\bf r}')\delta(\tau),\\
&&\Sigma({\bf r},{\bf r}'| \tau)=\Sigma^{H}({\bf r})\delta({\bf r}-{\bf r}')\delta(\tau)+\Sigma^{xc}({\bf r}, {\bf r}' | \tau),\\
&&\Sigma^{H}({\bf r}) = \int d{\bf r}_1\frac{\rho({\bf r}_1)}{|{\bf r}-{\bf r}_1|},\\
&&\rho({\bf r}) = G({\bf r}, {\bf r} | \tau \rightarrow 0^-).
\end{eqnarray}
Here, $\Sigma$ is the electron self-energy that captures the complicated electron-electron interactions. It is composed of the Hartree ($\Sigma^{H}$) and exchange-correlation ($\Sigma^{xc}$) components of the self-energy. $\Sigma^{H}$ relates to the updated electronic charge density ($\rho$) and $\Sigma^0$ is the sum of Hartree and exchange-correlation potentials in the reference K-S system. $\delta(\tau)$ is the Dirac delta function.

Given the high computational cost of calculating $\Sigma^{xc}$, the standard method used to find this quantity is the GW approximation, which can be expressed in real-space and Matsubara-time as\cite{Mahan2013}
\begin{equation}
\Sigma^{xc}({\bf r}, {\bf r}'|\tau)=-G({\bf r}, {\bf r}'|\tau)\cdot W({\bf r}, {\bf r}'|\tau).\label{gw1}
\end{equation}
Here, $W$ is the dynamically screened Coulomb potential, which describes the interactions between quasiparticles while including screening effects.  The screened Coulomb potential obeys the Dyson equation that reads
\begin{eqnarray}
W({\bf r}, {\bf r}'| \tau)&&= v({\bf r}, {\bf r}')\delta(\tau)+\int^{\beta}_{0}d\tau'\int d{\bf r}_1\int d{\bf r}_2 v({\bf r}, {\bf r}_1)\label{scrv}\\\nonumber
&& \times P({\bf r}_1, {\bf r}_2 | \tau-\tau')W({\bf r}_2, {\bf r}' |\tau'),
\end{eqnarray}
where $v({\bf r}, {\bf r}')=1/|{\bf r}-{\bf r}'|$ is the bare Coulomb potential, and $P$ is the irreducible polarization within RPA,
\begin{equation}
P({\bf r}, {\bf r}'| \tau)=G({\bf r}, {\bf r}'| \tau)\cdot G({\bf r}, {\bf r}'| -\tau).\label{prr}
\end{equation}

In addition, $\Sigma^{xc}(\tau)$ is often expressed as the sum of the exchange self-energy, $\Sigma^x(\tau)=-G(\tau)\cdot v\delta(\tau)$, which corresponds to the Fock exchange term, and the correlation self-energy, $\Sigma^c=-G(\tau)\cdot[W(\tau)-v\delta(\tau)]$. Note that the self-energy in Matsubara-time domain is simply a product of the Green's function and screened Coulomb potential, in contrast to the corresponding expression in Matsubara-frequency domain that requires a convolution of $G$ and $W$. The electron self-energy within the GW approximation (its exchange-correlation part is given in Eq.~(\ref{gw1})) correlates with the Green's function and thus both need to be solved self-consistently via Eq.~(\ref{dysonGF}).

The set of inter-correlated equations presented above allows us to compute $G$ and $\Sigma$ self-consistently. Once they are converged to the required accuracy, a Fourier transform of $\Sigma$ from the Matsubara-time to Matsubara-frequency domain is performed, i.e. $\{\Sigma(\tau)\}\rightarrow\{\Sigma(i\omega_n)\}$ where $\{\omega_n=(2n+1)\pi/\beta\}$ are the Matsubara frequencies with $n$ being integers, and the spatial dependence of $\Sigma$ is neglected for simplicity. This is then followed by an analytic continuation to real frequency space, $\{\Sigma(i\omega_n)\}\rightarrow\{\Sigma(\omega+i\eta)\}$, with $\eta$ being a positive infinitesimal number, which yields the Green's function in the real-frequency domain and the excitation spectrum of the system. 

\section{Implementation of the Self-consistent GW Method\label{dev}}
In this section, we describe the self-consistent GW approach in the Matsubara-time domain, which has been implemented in the modified version of the Elk FP-LAPW package.\cite{excitingplus, Kozhevnikov2010} The approach is essentially similar to the one proposed by Ku and Eguiluz,\cite{Ku2002} but with more efficient computational schemes. In particular, (i) we have employed the more efficient uniform power mesh (UPM) in Matsubara-time domain as proposed by Stan \textit{et al.},\cite{Stan2009} (ii) we have adopted the CPE for analytic continuation in conjunction with our scGW method, and (iii) full crystal symmetry has been taken into account to significantly reduce the computational load. We briefly summarize these improvements in the subsections below.

\subsection{Matsubara-time sampling}
The Green's function $G$ in the Matsubara-time domain varies smoothly in the range $0\le\tau\le\beta$ and does not have any singularity points, however it varies rapidly near $\tau=0$ and $\beta$. To capture this behavior without losing computational efficiency, we employ the UPM to sample the $\tau$-axis on the grid $\{\tau_0=0, \tau_1, \tau_2,...,\tau_M=\beta\}$ as proposed by Stan \textit{et al.},\cite{Stan2009} which is a modified version of the original one by Ku and Eguiluz.\cite{Ku2002} The UPM grid can be characterized by a pair of integers ($p$, $m$) as well as the length of the interval $\beta$, in which $p$ is the number of non-uniform sub-intervals generated between 0 and $\beta$ with $2m-1$ evenly distributed grid points inside each of these sub-intervals. A UPM mesh with given ($p$, $m$) results in 2$pm$+1 grid points (including the end points) in the interval. In this scheme, the grid density increases for values of $\tau$ closer to the end points in order to capture the varying behavior of $G$. Using this scheme, explicit evaluation of quantities such as the self-energy and Green's function, which is normally computationally expensive, now only requires a coarse UPM grid.  Thus, implementation of this grid significantly reduces the computational effort. For $\tau$ domain integrals that require knowledge of the integrand on a dense uniform $\tau$ grid, e.g. solving the Dyson equation, a higher-order interpolation such as cubic spline can be subsequently applied. 

\subsection{Scheme for scGW in Matsubara-time domain}
In this work, we expand and compute the Green's function $G$ and self-energy $\Sigma$ using the K-S basis (\{$\phi_{n{\bf k}}$\}), whereas we evaluate the polarization function $P$ and the screened Coulomb potential $W$ in reciprocal space (\{{\bf G}\}). We also assume that the quasiparticle wavefunctions are very similar to K-S eigenfunctions so that $\Sigma$ and $G$ become approximately diagonal in the K-S basis, significantly reducing computational effort. This approximation has been shown to provide reasonable results for a variety of systems.\cite{Shishkin2007, Usuda2002,Gomez2008} A direct generalization to include off-diagonal elements of $\Sigma$ is straightforward and will be completed in the future. The scGW approach is outlined below.

\subsubsection{Green's function in the reference K-S system $G^0$}
As a first step in scGW, we construct the Green's function in the reference K-S system ($G^0$):

\begin{equation}
G^{0}_{j}({\bf k}| \tau) = -\exp(-\epsilon_{j{\bf k}}\tau)[1 - n_F(\epsilon_{j{\bf k}})], \ \ \ \ \ 0 \le \tau \le \beta
\end{equation}
where $\{\epsilon_{j{\bf k}}\}$ are the K-S eigenenergies measured from the chemical potential $\mu$ of the system, $n_F = [\exp(\beta\epsilon_{j{\bf k}})+1]^{-1}$ is the Fermi-Dirac distribution, and $\bf k$ is a wave vector. In the zero-temperature limit, the results for a system with a non-zero band gap are insensitive to the choice of $\mu$ provided that it is placed inside the gap. 

\subsubsection{Irreducible polarization}

The irreducible polarization $P$ in the reciprocal space $\{{\bf G}\}$ can be obtained via Fourier transformations in Eq.~(\ref{prr}) that reads,
\begin{eqnarray}
P_{{\bf G}{\bf G}'}({\bf q}|\tau)=\frac{1}{\Omega}\int d{\bf r}\int d{\bf r}'e^{-i({\bf k}+{\bf G})\cdot{\bf r}}P({\bf r}, {\bf r}'|\tau)e^{i({\bf k}+{\bf G}')\cdot{\bf r}'},
\end{eqnarray}
where $\Omega$ is the volume of the unit cell, and ${\bf q}$ falling within the first Brillouin Zone (BZ). Using the relation between $P$ and $G$ in real space via Eq.~(\ref{prr}), and by transforming the Green's function from the Bloch-basis to real space,
\begin{eqnarray}
G({{\bf r},{\bf r}'}|\tau)=\sum^{BZ}_{{\bf k}}\sum_{j}\phi_{j{\bf k}}({\bf r})G_j({{\bf k}}|\tau)[\phi_{j{\bf k}}({\bf r}')]^*,
\end{eqnarray}
it is straight-forward to show that the irreducible polarization in the reciprocal space can be expressed as follows,
\begin{eqnarray}
&& P_{{\bf G}{\bf G}'}({\bf q}|\tau) = \frac{1}{\Omega N_{{\bf k}}}\sum_{\sigma}\sum^{BZ}_{{\bf k}}\sum_{j_1,j_2}M^{{\bf k}}_{j_2j_1}({\bf G},{\bf q})Q_{j_1j_2}({\bf k}, {\bf q}|\tau)[M^{\bf k}_{j_2j_1}({\bf G}',{\bf q})]^{*},\label{irpol} \\
&& Q_{j_1j_2}({\bf k}, {\bf q}|\tau)=G_{j_1}({\bf k}+{\bf q}|\tau)G_{j_2}({\bf k}|-\tau),\nonumber\\
&& M^{{\bf k}}_{nm}({\bf G}, {\bf q})=\sum_{\sigma}\int d{\bf r}[\psi^{\sigma}_{n{\bf k}}({\bf r})]^{*}e^{-i({\bf q}+{\bf G})\cdot{\bf r}}\psi^{\sigma}_{m{\bf k}+{\bf q}}({\bf r})\label{mmatrix}.
\end{eqnarray}
Here, $j_1$ and $j_2$ are dummy band indices that run through both valence and conduction bands, $\sigma$ is the dummy spin index, ${\bf q}$ is a reciprocal vector, and ${\bf G}$ is a reciprocal lattice vector. It is clear that the irreducible polarization $P$ at any two distinct $\tau_1$ and $\tau_2$ in [0, $\beta$] are decoupled. Therefore, parallelization over $\tau$ can be performed efficiently when $P$ is evaluated.

\subsubsection{Screened Coulomb potential}
The screened Coulomb potential ($W$) can be computed once $P$ is determined. Instead of directly solving for $W$, during which the emergence of the Dirac delta function $\delta(\tau)$ (see Eq.~(\ref{scrv})) may lead to numerical instability, we work with $\tilde{W}(\tau)\equiv W(\tau) - v\delta(\tau)$ (only $\tau$ dependence is indicated for simplicity). This formulation yields a correlation self-energy, $\Sigma^{c}(\tau)=-G(\tau)\cdot\tilde{W}(\tau)$, and exchange self-energy, $\Sigma^{x}(\tau)=-G(\tau)v\cdot\delta(\tau)$, such that $\Sigma^{xc}(\tau)=\Sigma^{x}(\tau)+\Sigma^{c}(\tau)$. In reciprocal space and Matsubara-time domain, $\tilde{W}$ obeys the following Dyson equation
\begin{eqnarray}
\tilde{W}_{{\bf G}{\bf G}'}({\bf q}|\tau)=&& \sum_{{\bf G}_2}\left[\sum_{{\bf G}_1}v_{{\bf G}{\bf G}_1}({\bf q})P_{{\bf G}_1{\bf G}_2}({\bf q}|\tau)\right]v_{{\bf G}_2{\bf G}'}({\bf q})\nonumber\\
&& +\int^{\beta}_0 d\tau'\sum_{{\bf G}_2}\left[\sum_{{\bf G}_1}v_{{\bf G}{\bf G}_1}({\bf q})P_{{\bf G}_1{\bf G}_2}({\bf q}|\tau-\tau')\right]\tilde{W}_{{\bf G}_2{\bf G}'}({\bf q}|\tau'),
\end{eqnarray}
where $v_{{\bf G}{\bf G}'}({\bf q})=4\pi\delta_{{\bf G}{\bf G}'}/|{\bf q}+{\bf G}|^2$ is the Fourier transform of the bare Coulomb potential. We follow the algorithm proposed by Stan \textit{et al.}\cite{Stan2009} to discretize the $\tau$-axis using the generated UPM grid. The above equation can then be re-arranged to form a linear matrix equation that reads
\begin{eqnarray}
&& \sum_{r=0}^{M}\sum_{{\bf G}_2}\left[\delta_{{\bf G}{\bf G}_2}\delta_{p,r}-A_{{\bf G}{\bf G}_2}({\bf q}|\tau^{(p)}-\tau^{(r)})\Delta\tau^{(r)}\right]\tilde{W}_{{\bf G}_2{\bf G}'}({\bf q}|\tau^{(r)})\nonumber\\
&& =\sum_{{\bf G}_2}A_{{\bf G}{\bf G}_2}({\bf q}|\tau^{(p)})v_{{\bf G}_2{\bf G}'}({\bf q}),\label{tildeW}\\
&& A_{{\bf G}{\bf G}_2}({\bf q}|\tau)\equiv\sum_{{\bf G}_1}v_{{\bf G}{\bf G}_1}({\bf q})P_{{\bf G}_1{\bf G}_2}({\bf q}|\tau).\nonumber
\end{eqnarray}
Here, the increments $\Delta\tau$ are positive, with $\Delta\tau^{(i)}=(\tau^{i+1}-\tau^{i-1})/2$ for $1\le i\le M-1$. At the end points, $\Delta\tau^{(0)}=(\tau^{1}-\tau^{0})/2$ and $\Delta\tau^{(M)}=(\tau^{M}-\tau^{M-1})/2$. 

\subsubsection{Evaluating the self-energy}
With $\tilde{W}(\tau)$ and $G(\tau)$ in hand, the correlation self-energy ($\Sigma$) can be evaluated as 
\begin{eqnarray}
&& \Sigma^{c}_n({\bf k}|\tau)=-\frac{1}{\Omega N_{{\bf k}}}\sum^{BZ}_{{\bf q}}\sum_{{\bf G}{\bf G}'}\sum_{j}\left[ M^{{\bf k}-{\bf q}}_{jn}({\bf G},{\bf q})\right]^{*}O^{{\bf G}{\bf G}'}_j({\bf k},{\bf q}|\tau)M^{{\bf k}-{\bf q}}_{jn}({\bf G}',{\bf q}),\label{sigc}\\
&& O^{{\bf G}{\bf G}'}_j({\bf k}-{\bf q}|\tau)=G_{j}({\bf k}-{\bf q}|\tau)\tilde{W}({\bf q}|\tau).\nonumber
\end{eqnarray}
On the other hand, the exchange self-energy $\Sigma^{x}$ is evaluated in real-space due to the slow convergence of $\Sigma^{x}$ in reciprocal space,\cite{Chu2014}
\begin{eqnarray}
\Sigma^{x}_{n{\bf k}}=-\sum_{{\bf k}'\in BZ}\sum^{occ}_{m}\int d{\bf r}\sum_{\sigma}[\psi^{\sigma}_{n{\bf k}}({\bf r})]^{*}\psi^{\sigma}_{m{\bf k}'}({\bf r})\int d{\bf r}'\frac{\sum_{\sigma}'[\psi^{\sigma'}_{m{\bf k}'}({\bf r}')]^{*}\psi^{\sigma'}_{n{\bf k}}({\bf r}')}{|{\bf r}-{\bf r}'|}f_{m{\bf k}'},\label{sigx}
\end{eqnarray}
where $f_{j{\bf k}}=G_{j}({\bf k}|0^{-})$ is the occupation number of the K-S eigenfunction in spinor form, $\Psi_{j{\bf k}'}({\bf r})=[\psi^{\uparrow}_{j{\bf k}}({\bf r}),\psi^{\downarrow}_{j{\bf k}}({\bf r})]$. Similarly, the Hartree potential is expressed as 
\begin{eqnarray}
\Sigma^{H}_{n{\bf k}}=\sum_{\sigma}\int d{\bf r}|\psi^{\sigma}_{n{\bf k}}({\bf r})|^2\int d{\bf r}'\frac{\sum_{{\bf k}'\in BZ}\sum_{\sigma',m}|\psi^{\sigma}_{m{\bf k}'}({\bf r}')|^2}{|{\bf r}-{\bf r}'|}f_{m{\bf k}'}.\label{sigh}
\end{eqnarray}

\subsubsection{Dressed Green's function}
During the scGW calculation,  the Green's function ($G$) is updated in each iteration using the newly obtained self-energy $\Sigma$ in the Dyson equation, which reads
\begin{eqnarray}
&& G^{N}_{j}({\bf k}|\tau) = G^{0}_{j}({\bf k}|\tau)+\int^{\beta}_{0}d\tau_2 Z_{j{\bf k}}(\tau,\tau_2)G^{N}_{j}({\bf k}|\tau_2),\label{dysongf}\\
&& Z_{j{\bf k}}(\tau, \tau_2) = Z^{x}_{j{\bf k}}(\tau, \tau_2) + Z^{c}_{j{\bf k}}(\tau, \tau_2), \\
&& Z^{x}_{j{\bf k}}(\tau, \tau_2)=G^{0}_{j}({\bf k}|\tau-\tau_2)[\Sigma^{x}_{N,j}({\bf k})+\Sigma^{H}_{N,j}({\bf k})-\Sigma_{0,j}({\bf k})],\label{zxjk}\\
&& Z^{c}_{j{\bf k}}(\tau, \tau_2)=\int^{\beta}_0d\tau_1 G^{0}_{j}({\bf k}|\tau-\tau_1)\cdot\Sigma^{c}_{N,j}({\bf k}|\tau_1-\tau_2)\label{zcjk}.
\end{eqnarray}

The integrals along the $\tau$ axis in Eqs.~(\ref{dysongf}) and (\ref{zcjk}) may have substantial numerical errors when performed on the UPM mesh that becomes coarse farther away from the end points of $0\le\tau\le\beta$. To overcome this issue, a cubic spline interpolation is applied to the Green's function and self-energy elements between two adjacent $\tau$ grid points, in which the increment $\Delta\tau$ in the resulting dense uniform $\tau$ grid is selected as $\tau_1-\tau_0$. This is also the smallest $\Delta\tau$ in the UPM mesh. Then the Dyson equation is solved on the generated, denser uniform $\tau$ mesh. Similar to the algorithm for $\tilde{W}$ as proposed by Stan \textit{et al.},\cite{Stan2009} the Dyson equation for $G$ along $\tau$ axis can be re-arranged to form a linear matrix equation. 
\begin{eqnarray}
\sum^{N}_{r=1}[\delta_{p,r}-\Delta\tau^{(r)}Z_{j{\bf k}}(\tau^{(p)},\tau^{(r)})]G^{N}_{j}({\bf k}|\tau^{(r)})=G^{0}_{j}({\bf k}|\tau^{(p)}).\label{dysong}
\end{eqnarray}

During the scGW calculation, we repeat the steps mentioned above in each iteration using the newly obtained Green's function $G$, as indicated in Eqs.(\ref{irpol}), (\ref{tildeW})-(\ref{sigh}) and (\ref{zxjk})-(\ref{dysong}). We solve for the self-energy and the Green's function in Matsubara-time domain self-consistently until any given accuracy is reached. Note that this corresponds to the single-shot G$_{0}$W$_0$ if the self-consistent calculation is terminated at the first iteration. The approximated calculation known as GW$_0$ can also be performed if the screened Coulomb potential $W$ is kept constant after the first iteration whereas $G$ is updated during the self-consistent loop.

\subsubsection{Analytic continuation}
To obtain quantities that can be measured in experiments, such as the excitation spectrum, knowledge of $G$ and $\Sigma$ in the real-frequency domain is required. This is achieved by a two-step procedure performed after calculating the converged self-energy in the Matsubara-time domain ($\Sigma^{xc}(\tau)$).  First, a Fourier transformation from Matsubara-time to Matsubara-frequency domain is employed. For a given band index $n$ and ${\bf k}$, this reads
\begin{eqnarray}
\Sigma^{xc}_{j}({\bf k}|i\omega_n)=\int^{\beta}_0 d\tau e^{i\omega_n\tau}\Sigma^{xc}_j({\bf k}|\tau)\label{trans},
\end{eqnarray}
where $\omega_n=(2n+1)\pi/\beta$ is the Matsubara frequency with $n$ being an integer. We use a cubic spline interpolation of the UPM grid for the accurate evaluation of the integral. Second, we implement analytic continuation using the CPE method proposed by Staar \textit{et al.}\cite{Staar2014} to yield the self-energy in the real-frequency domain ($\Sigma^{xc}(\omega+i\eta)$).  Unlike the commonly used Pade approximation,\cite{Vidberg1977} where the self-energy elements are simply expanded as polynomials, the CPE takes advantage of the fact that the self-energy in the upper complex plane ($z$) can be expressed as
\begin{eqnarray}
&& \Sigma_j^{xc}({\bf k},z)=\frac{1}{2\pi}\int^{+\infty}_{-\infty}d\omega\frac{\operatorname{Im}[\Sigma^{xc}_{j}({\bf k}, \omega+i\eta)]}{\omega-z},\\
&& \operatorname{Im}[\Sigma^{xc}_j({\bf k},\omega+i\eta)]<0,\label{nesig}
\end{eqnarray}
where $\eta$ is a positive infinitesimal and Eq.~(\ref{nesig}) arises from causality. For each $j$ and ${\bf k}$, $\operatorname{Im}\Sigma^{xc}_j({\bf k}|\omega+i\eta)$ can be expanded as a set of piecewise linear functions of $\omega$ with undetermined coefficients $\{a_{nj}({\bf k})\}$. This leads to $\Sigma^{xc}_j({\bf k},z)=\sum_{m}a_{mj}({\bf k})\Phi_{mj,{\bf k}}(z)$ where $\Phi_{mj,{\bf k}}(z)$ is some analytic function defined in the upper complex plane $z$. With the set of computed elements $\{\tilde{\Sigma}^{xc}_j({\bf k}|i\omega_m)\}$ in hand and Eq.~(\ref{nesig}) as the constraint, for each given $j$ and ${\bf k}$, $\{a_{nj}({\bf k})\}$ are then determined by minimizing the norm function $\Omega$ defined as
\begin{eqnarray}
\Omega_{j}({\bf k})=\sum^{M}_{m=0}|\tilde{\Sigma}^{xc}_{j}({\bf k}| i\omega_{m})-\Sigma^{xc}_{j}({\bf k}|i\omega_m)|^2,
\end{eqnarray}
with $M$ being the number of positive Matsubara frequencies. Given the fitted $\Sigma_{j}^{xc}({\bf k}|\omega)$, for each $j$ and ${\bf k}$, the Green's function associated with the interacting system can be determined using
\begin{eqnarray}
G_{j}({\bf k}|\omega)=\frac{1}{[G^0_j({\bf k}|\omega)^{-1}-\Sigma^{xc}_j({\bf k}|\omega)-\Sigma^{H}_j({\bf k})]}.
\end{eqnarray}
The quasiparticle energies, and hence the electronic band gap can be directly obtained from the spectral function $A_{j{\bf k}}(\omega)=-\frac{1}{\pi}$Im$[G_{j{\bf k}}(\omega)]$ for given $j$ and ${\bf k}$.

\subsection{Use of Crystal Symmetry for Computational Speedup}
Calculating the elements of $\Sigma^{c}(\tau)$ can be computationally expensive as it involves the evaluation of Eqs.~(\ref{irpol}), (\ref{mmatrix}), (\ref{tildeW}) and (\ref{sigc}). Such computational effort can be considerably reduced using crystal symmetry to decrease the number of required operations. The allowed crystal symmetry operations are those that leave the Hamiltonian invariant. Using these operations, reciprocal vectors in the first BZ $\{{\bf k}_{BZ}\}$ are decomposed to a number of subsets. The reciprocal vectors in each of these subsets are related via the action of the symmetry operations. Therefore, the first BZ can be represented using a reduced set of ${\bf k}$ vectors that form the irreducible BZ, denoted as $\{{\bf k}_{IBZ}\}$.

Suppose $S_u\equiv\{({\bf R}_i|{\bf t}_i), i=1,...,N_u\}$ is the set of symmetry operations in which ${\bf R}$ is a $3\times3$ rotation matrix and ${\bf t}$ the translation vector in real space. The application of a given symmetry operation, $B_i=({\bf R}_i|{\bf t}_i)$, on the real-space vector ${\bf r}$ and reciprocal vector in IBZ lead respectively to
\begin{eqnarray}
&& B_i{\bf r}={\bf R}_i{\bf r}+{\bf t}_i\label{rso},\\
&& {\bf k}_{BZ}=B_i{\bf k}_{IBZ}={\bf R}_i{\bf k}_{IBZ}+{\bf G}_{{\bf R}i},\label{qibz2bz}
\end{eqnarray}
where ${\bf G}_{{\bf R}i}$ is the reciprocal lattice vector that brings ${\bf R}_i{\bf k}_{IBZ}$ back to the 1st BZ. For a given ${\bf q}_{BZ}$ that is associated with ${\bf q}_{IBZ}$ via ${\bf R}$ and ${\bf G}$ using Eq.~(\ref{qibz2bz}), it is straight forward to prove that the plane-wave matrix $M$ in Eq.(\ref{mmatrix}), the irreducible polarization $P(\tau)$ in Eq.(\ref{irpol}), and $\tilde{W}(\tau)$ in Eq.(\ref{tildeW}) obey the following relations
\begin{eqnarray}
&& M^{{\bf k}}_{nm}({\bf G}, {\bf q}_{BZ})=M^{{\bf R}^{-1}{\bf k}}_{nm}[{\bf G}_1, {\bf q}_{IBZ}]\exp[-i({\bf R}{\bf q}_{IBZ}+{\bf G}_{{\bf R}}+{\bf G})\cdot{\bf t}],\\
&& P_{{\bf G}{\bf G}'}({\bf q}_{BZ}|\tau)=P_{{\bf G}_1{\bf G}'_1}({\bf q}_{IBZ}|\tau)\exp[-i({\bf G}-{\bf G}')\cdot{\bf t}],\\
&& \tilde{W}_{{\bf G}{\bf G}'}({\bf q}_{BZ}|\tau)=\tilde{W}_{{\bf G}_1{\bf G}'_1}({\bf q}_{IBZ}|\tau)\exp[-i({\bf G}-{\bf G}')\cdot{\bf t}],
\end{eqnarray}
where ${\bf G}_1={\bf R}^{-1}({\bf G}+{\bf G}_{{\bf R}})$ and ${\bf G}'_1={\bf R}^{-1}({\bf G}'+{\bf G}_{{\bf R}})$. It follows that the correlation self-energy can be re-arranged as
\begin{eqnarray}
\Sigma^{c}_n({\bf k}|\tau)=&&-\frac{1}{\Omega N_{{\bf k}}}\sum_{{\bf q}_{IBZ}}\sum_{{\bf R}}\sum_{{\bf G}{\bf G}'}\tilde{W}_{{\bf G}{\bf G}'}({\bf q}_{IBZ}|\tau)\sum_{j}[M^{{\bf R}^{-1}{\bf k}-{\bf q}_{IBZ}}_{jn}({\bf G},{\bf q}_{IBZ})]^{*}\nonumber\\
&& \times G_{j}({\bf R}^{-1}{\bf k}-{\bf q}_{IBZ}|\tau)M^{{\bf R}^{-1}{\bf k}-{\bf q}_{IBZ}}_{jn}({\bf G}',{\bf q}_{IBZ}).
\end{eqnarray}

Here, ${\bf R}^{-1}{\bf k}-{\bf q}_{IBZ}$ is assumed to fall in the set of $\{{\bf k}_{BZ}\}$ vectors. It is thus sufficient to compute the summands in the above equation for the sets of $\{{\bf q}_{IBZ}\}$ and $\{{\bf k}_{BZ}\}$ vectors, which leads to significant reduction to computational time. Similarly, the computation of the elements of $\Sigma^x$ can be sped up with the use of symmetry operations for ${\bf k}$. According to Eq.~(\ref{sigx}), in particular, ${\bf k}$ associated with $\Sigma^x$ can be confined to the IBZ, whereas ${\bf k}'$ runs over the 1st BZ.

\section{Computational Details}
The scGW scheme has been applied to calculate the electronic band gaps of 18 diverse semiconductors and insulators. We have adopted the experimental lattice parameters of 5.43 $\AA$ (Si), 5.658 $\AA$ (Ge), 5.66 $\AA$ (GaAs), 4.35 $\AA$ (SiC), 5.91 $\AA$ (CaSe), 3.57 $\AA$ (diamond), 5.64 $\AA$ (NaCl), 4.21 $\AA$ (MgO), 3.62 $\AA$ (cubic BN), 4.01 $\AA$ (LiF), 3.91 $\AA$ (cubic \ce{SrTiO3}), 4.27 $\AA$ (\ce{Cu2O}), 4.52 $\AA$ (GaN), 4.58 $\AA$ (zinc-blende ZnO), 5.42 $\AA$ (zinc-blende ZnS), 5.67 $\AA$ (zinc-blende ZnSe), 6.05 $\AA$ (zinc-blende CdSe) and 5.82 $\AA$ (zinc-blende CdS) throughout this work. All DFT calculations have been carried out using the modified version of the Elk FP-LAPW package.\cite{excitingplus, Kozhevnikov2010} The augmented plane wave + local orbitals (APW+lo) basis\cite{Sjostedt2000} with a single second-order local orbital per core or semi-core state has been adopted. The local density approximation (LDA)\cite{Perdew1992} has been utilized for the exchange-correlation functionals. When expanding the interstitial potential and charge density, the maximum length of the reciprocal lattice vector $|{\bf G}|$ has been chosen as 12 a.u. The angular momentum has been truncated as $\ell_{max}=8$ for the expansions of muffin-tin charge density, potential and wave function. In the expansion of the wave function, $|{\bf G}+{\bf k}|_{max}=8.0/R_{avg}$ has been used, where $R_{avg}$ is the average of the muffin-tin radii ($R^{MT}$) in each system. Linearization energy ($E_{\ell, \nu}$), which is associated with each radial function labeled with $\nu$, is chosen at the center of the corresponding band with $\ell$-like character. The first BZ has been sampled by a $4\times4\times4$ ${\bf k}$-mesh for all the systems except for diamond, where a $6\times6\times6$ ${\bf k}$ mesh has been used instead. All the aforementioned parameters have been carefully tested to achieve total energy convergence. 

In the GW calculations, the cutoff for $|{\bf G}+{\bf q}|$ used in Eqs.~(\ref{irpol}) and (\ref{sigh}) has been set 4.0 a.u. for all the systems except for the systems of ZnO, diamond and cubic BN (c-BN), where a cutoff of 5.0 a.u. has been selected instead. These length cutoffs correspond to a kinetic-energy cutoff of 16 Ry and 25 Ry, respectively. The Matsubara-time ($\tau$) domain has been sampled with a (9, 5) UPM mesh, which consists of 81 grid points between 0 and $\beta$ associated with an artificial temperature of 300 K. A minimum of 150 conduction bands have been included for the band summations in Eqs.~(\ref{irpol}) and (\ref{sigh}) for the systems studied to ensure the convergence of the band gaps. In the GW$_0$ and scGW calculations, states with an energy falling in the energy window of $\pm$ 15 eV around the DFT-LDA Fermi energy have been updated, and the number of iterations has been set 4. In the transformation indicated in Eq.~(\ref{trans}), a set of 128 positive Matsubara frequencies has been adopted, which is subsequently used in analytic continuation schemes of both CPE and Pade approximation. For comparison, we have also performed G$_0$W$_0$ calculations using the plasmon-pole approximation (PPA), in which we have selected the model proposed by Godby and Needs\cite{Godby1989} that has proven to be in consistent agreement with numerical integration method.\cite{Stankovski2011, Chu2014}  All the above parameters are carefully examined to ensure the band gap values converged to within 50 meV. 

\section{Results and Discussion\label{results}}

\subsection{Benchmarking Si, Ge and GaAs}
\begin{figure}[htp]
\begin{center}
\subfigure[]{\label{fig:gf_tau_v}
\includegraphics[width=0.6\textwidth]{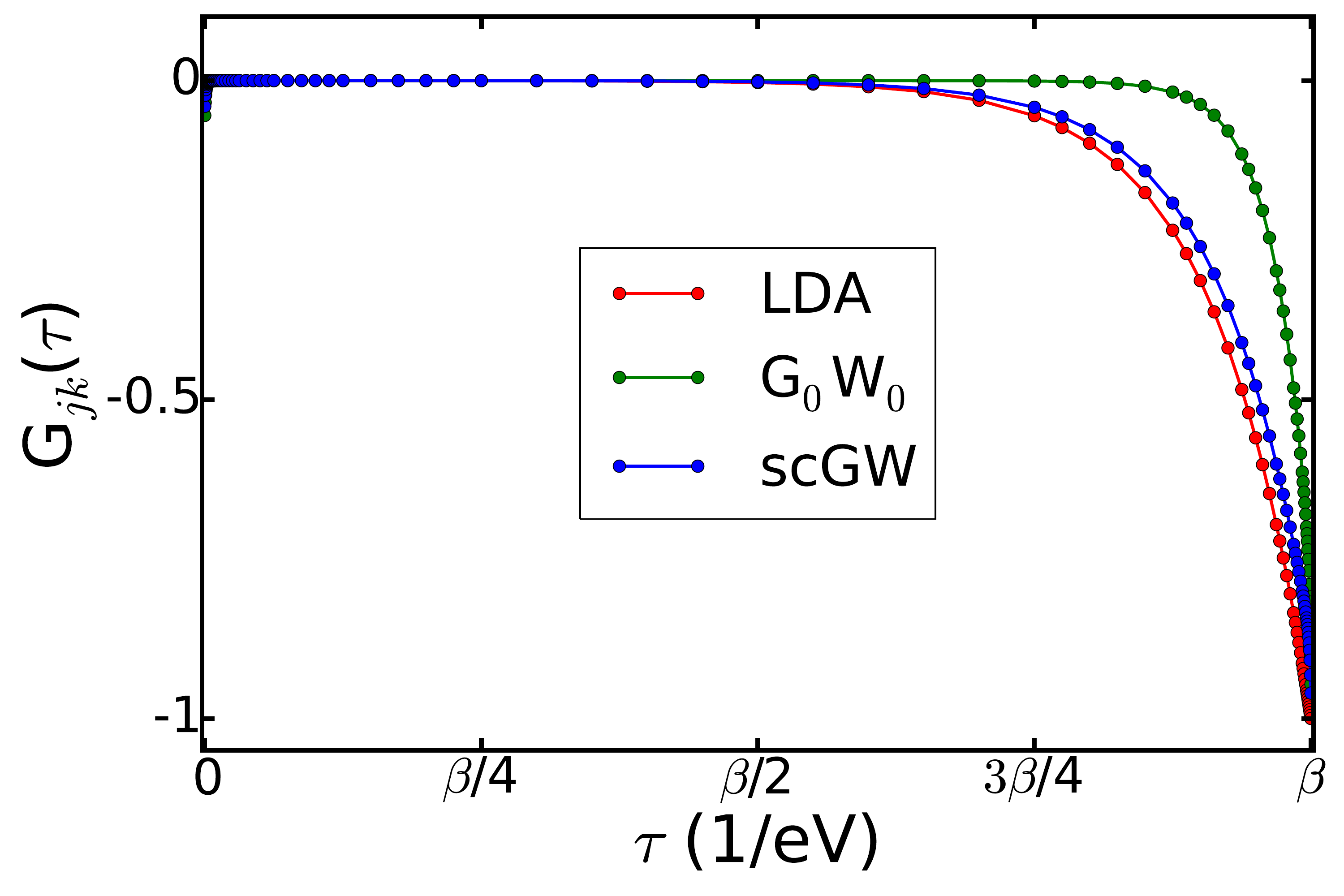}
}
\subfigure[]{\label{fig:gf_tau_c}
\includegraphics[width=0.6\textwidth]{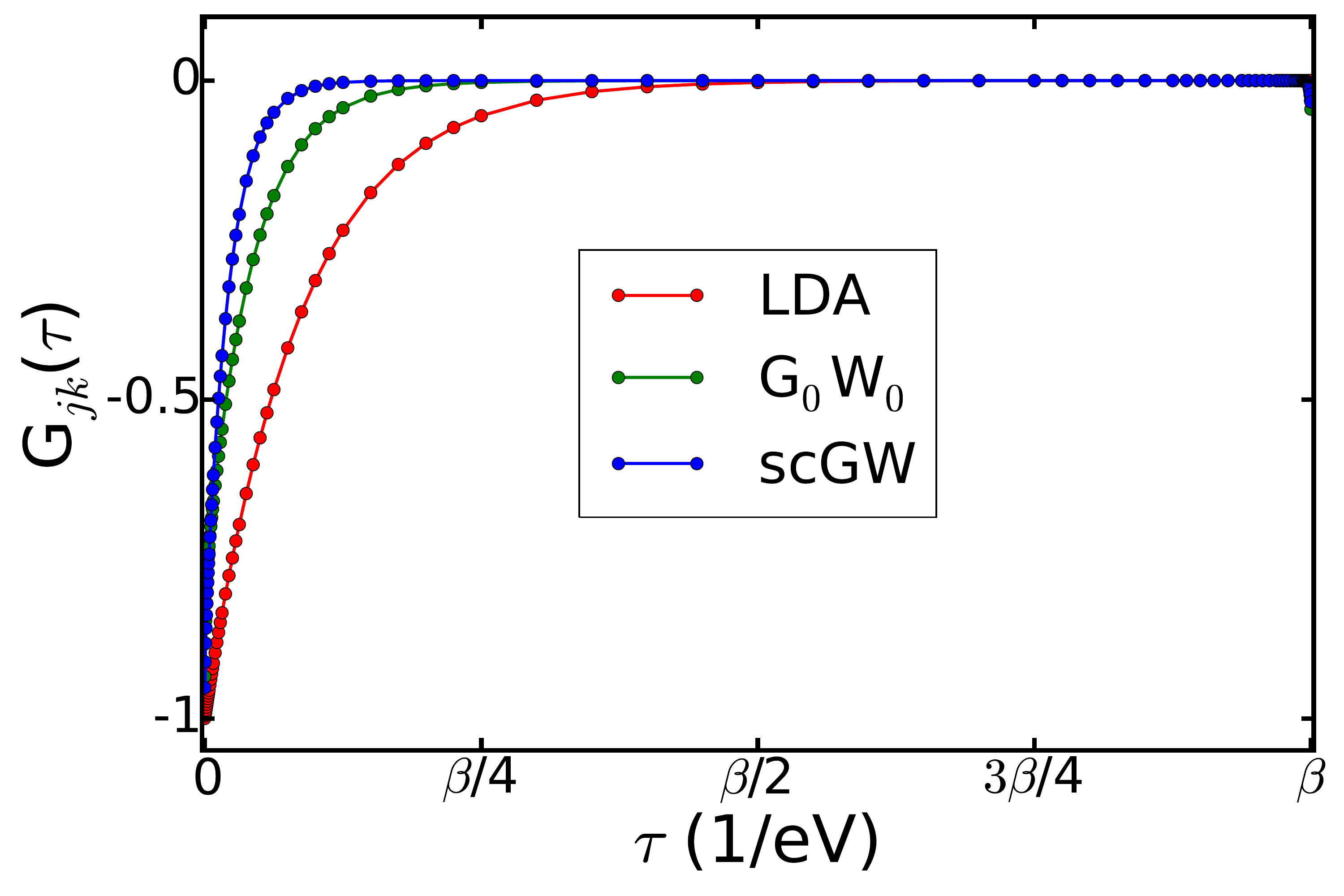}
}
\subfigure[]{\label{fig:gf_iwn}
\includegraphics[width=0.6\textwidth]{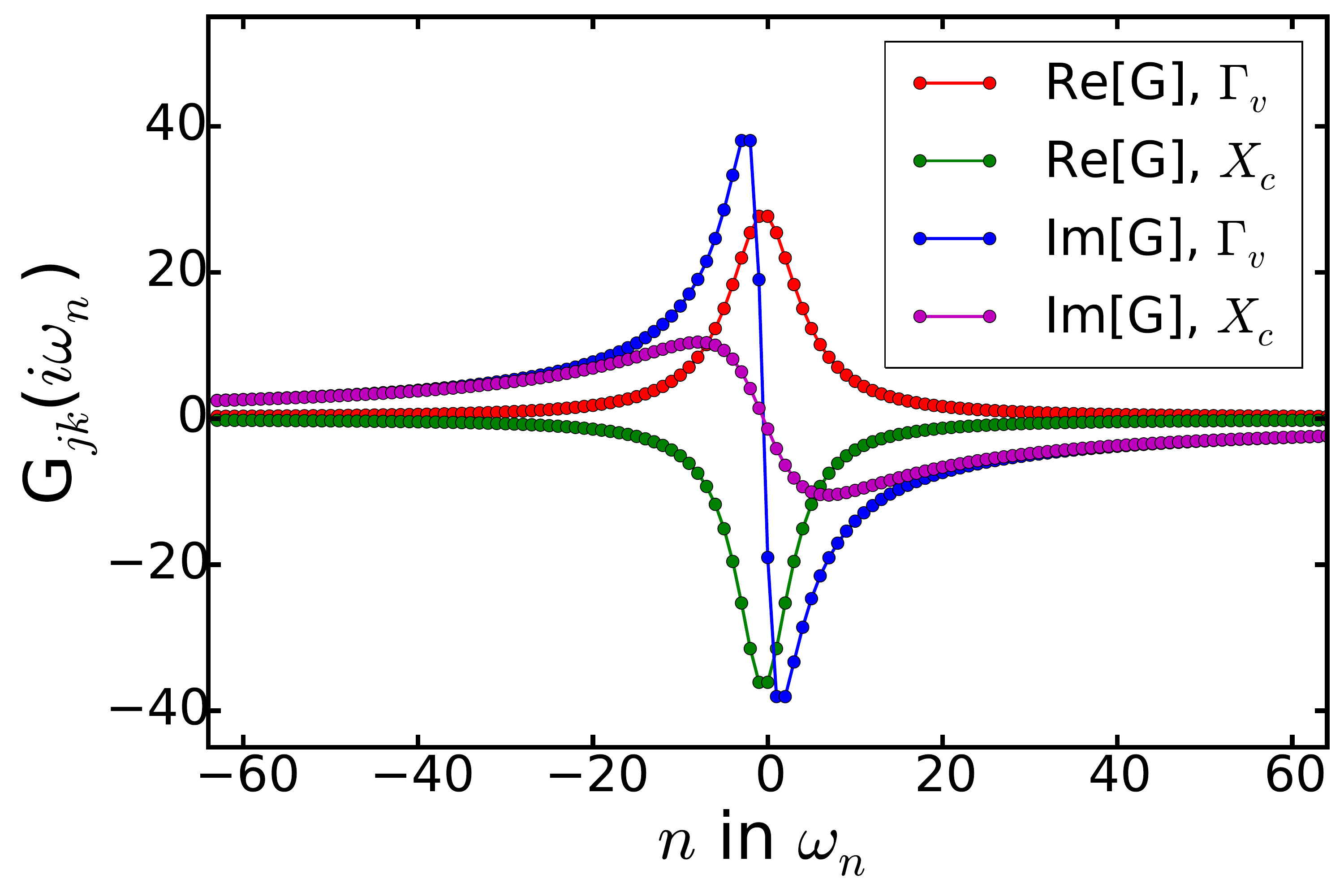}
}
\caption{(Color online) Single-particle Green's functions $G_{n{\bf k}}$ at the (a) valence band maximum ($\Gamma_v$) and (b) conduction band minimum ($X_c$) of bulk Si in Matsubara-time domain. (c) Single-particle Green's function of bulk Si in Matsubara-frequency domain from scGW calculations.}
\label{gf_plot}
\end{center}
\end{figure}

We first apply the Matsubara-time GW method to study the electronic properties of bulk silicon (Si), a prototypical system that has been studied as a benchmark for previous GW code developments. Figure~\ref{gf_plot}(a) and (b) illustrate the Matsubara-time Green's functions ($G(\tau)$) of the band-edge states at $\Gamma_v$ and $X_c$ at different levels of GW approximations, where $K_v$ ($K_c$) denotes the highest occupied (lowest unoccupied) single-particle state at $K$. $G(\tau)$ approaches -1 and 0 at each end of the $\tau$ axis. For the case of the valence (conduction) band state in a semiconductor/insulator, $G(\beta^-)\rightarrow -1 (0)$ to account for the occupation number of that state. It can be seen that in Matsubara-time domain, scGW leads to substantial changes of $G$ compared to those from G$_0$W$_0$. It is worth pointing out that the dressed $G$ at $\Gamma_v$ upon scGW becomes very similar to that from LDA, i.e. $G^0$. On the other hand, the scGW leads to more deviation of $G$ at $X_c$ from $G^0$, suggesting that GW corrections to the conduction bands are likely more pronounced than to the valence bands. Figure~\ref{gf_plot} (c) shows the typical Green's function in Matsubara-frequency domain (both real and imaginary parts) for the band edge states of bulk Si from scGW calculations. 

\begin{figure}[htp]
\begin{center}
\subfigure[]{\label{fig:aw_si1}
\includegraphics[width=0.48\textwidth]{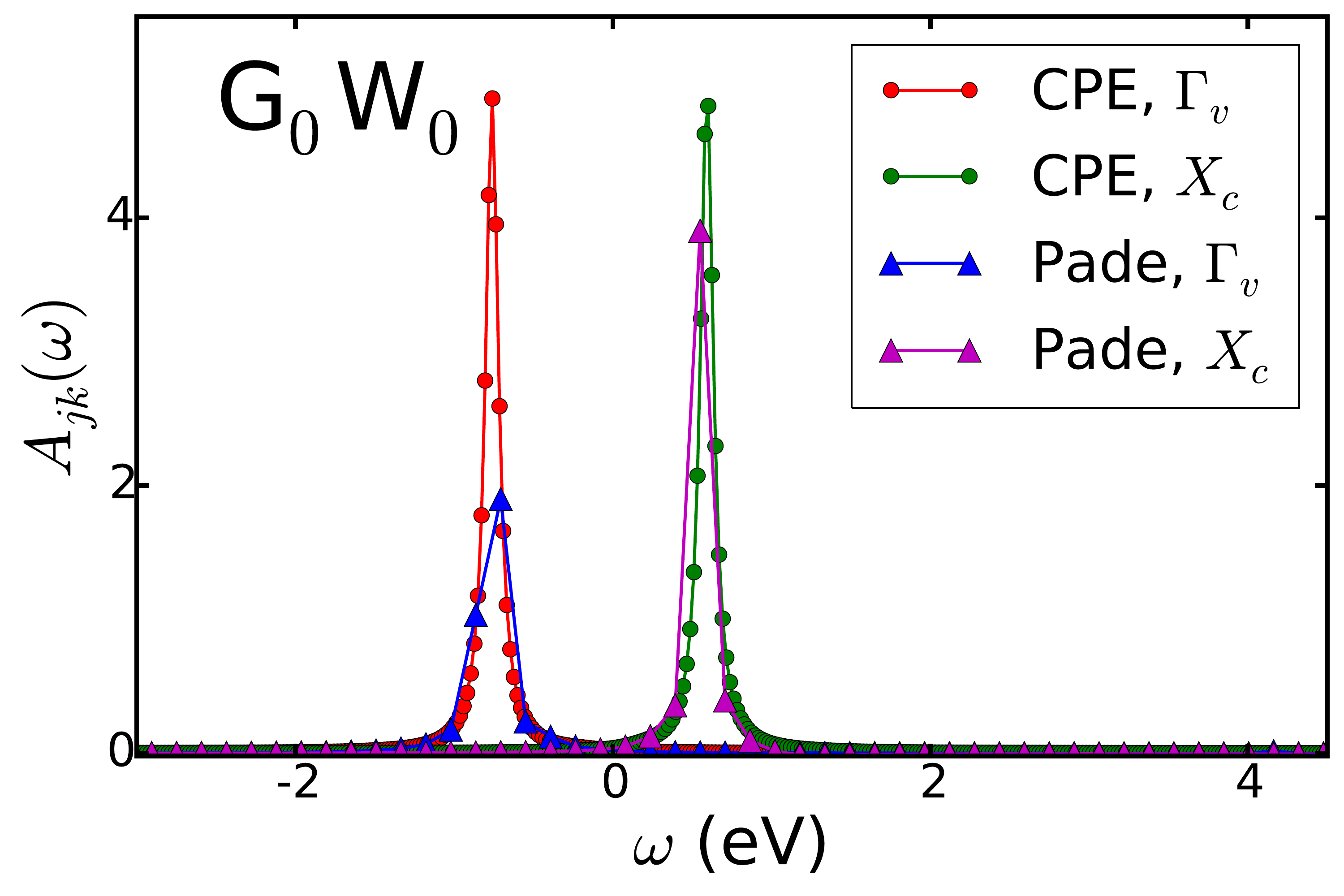}
}
\subfigure[]{\label{fig:aw_si2}
\includegraphics[width=0.48\textwidth]{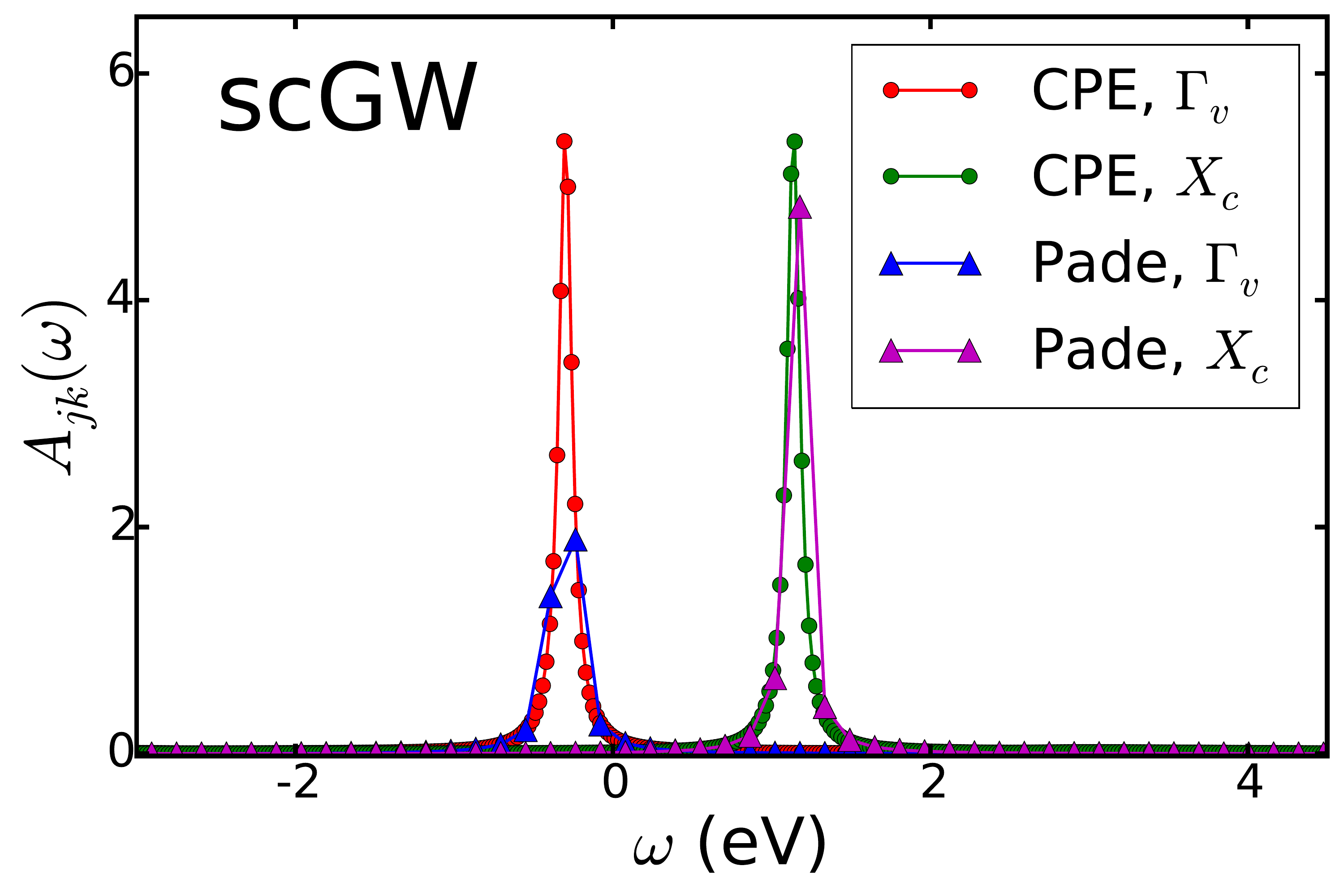}
}
\subfigure[]{\label{fig:aw_sto1}
\includegraphics[width=0.48\textwidth]{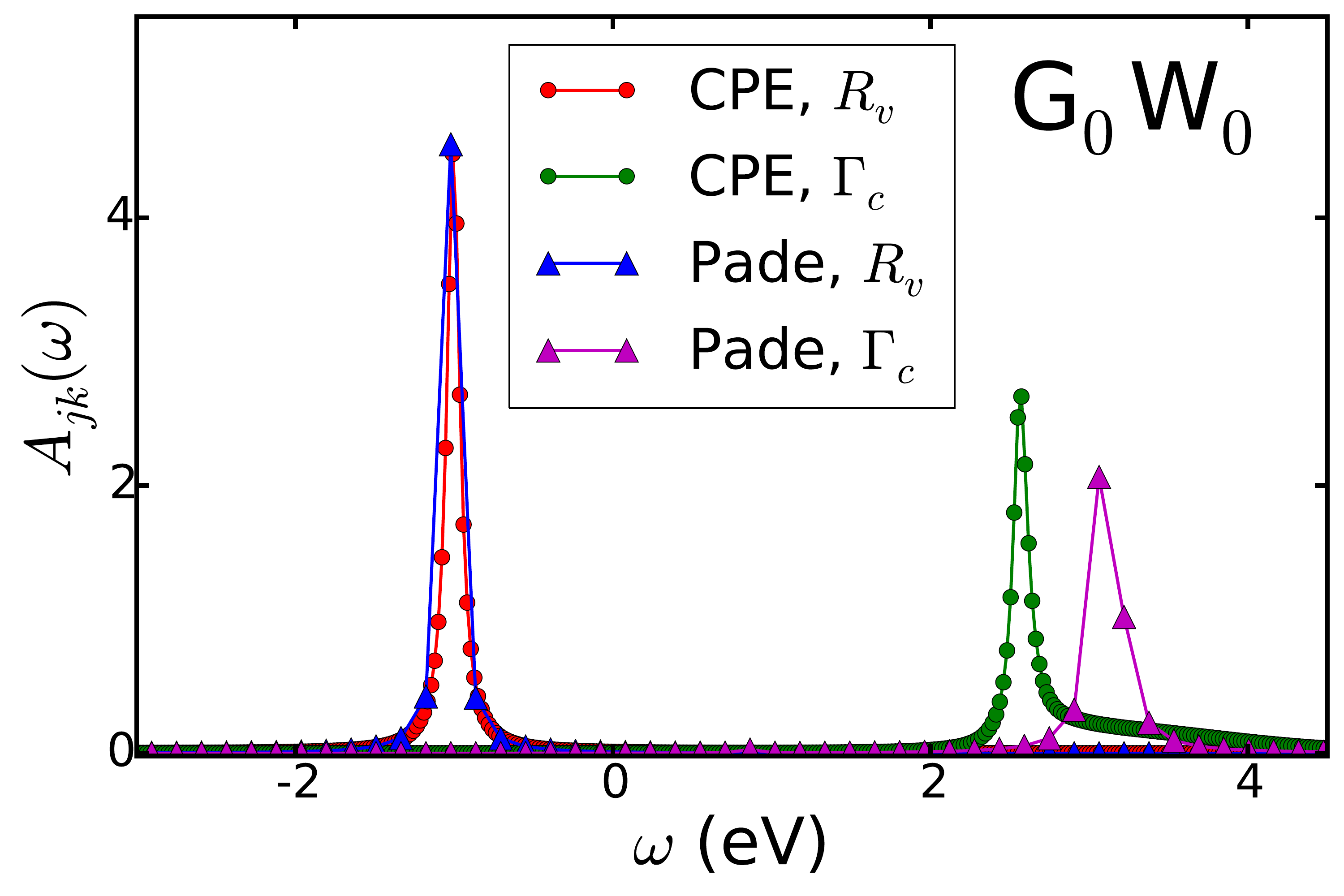}
}
\subfigure[]{\label{fig:aw_sto2}
\includegraphics[width=0.48\textwidth]{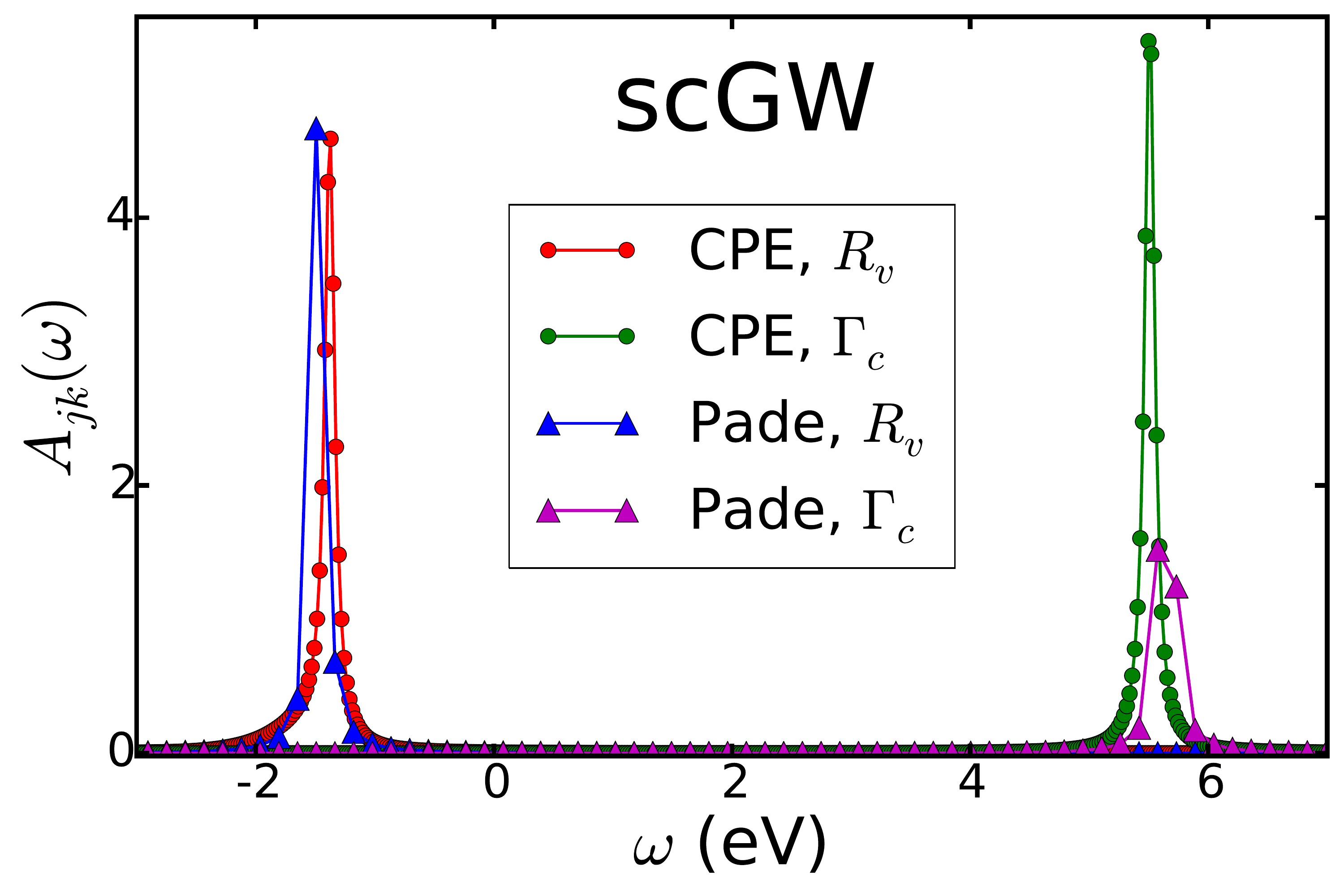}
}
\caption{(Color online) Spectral functions of band edge states $\Gamma_v$ and $X_c$ of bulk Si from (a) G$_0$W$_0$, and (b) scGW. Spectral functions of band edge states $R_v$ and $\Gamma_c$ of \ce{SrTiO3} from (c) G$_0$W$_0$, and (d) scGW.}
\label{aw_plot}
\end{center}
\end{figure}

The calculated band gaps for bulk Si are tabulated in Table~\ref{Sigaps}. When the non-self-consistent G$_0$W$_0$ calculation is performed, the direct band gap at $\Gamma$ and the indirect band gap from $\Gamma$ to $X$ are, respectively, 3.40 eV and 1.38 eV. These values are in relatively good agreement with experimental values.\cite{Hellwege1987} Compared to those obtained from plane-wave pseudopotential (PP)-based and/or all-electron G$_0$W$_0$, our computed G$_0$W$_0$ direct band gap and the indirect band gap are 0.1 eV and 0.2 eV higher, respectively. Moreover, we notice that band gap values are further increased by 0.2 eV upon implementing the partially self-consistent, GW$_0$ calculation. However, fully self-consistent GW brings the direct and indirect band gap values close to those calculated within the G$_0$W$_0$ approximation. Our scGW results are also comparable with the previous study by Ku \textit{et al.}, which uses a similar implementation to the present method. We also compare the results at different levels of GW using either the CPE method or Pade approximation. Band gap results  using the Pade approximation generally agree well with those using CPE analytic continuation within 0.02 eV.  However, the Si direct band gap value predicted by the Pade approximation is 0.16 eV higher than the CPE value, and is 0.09 eV higher than the value by Ku \textit{et al.}. This also shows that CPE results are generally in better agreement with experiment. In addition, all levels of GW calculations, from G$_0$W$_0$ to scGW, overestimate the experimental indirect band gap value by 0.13 to 0.34 eV.  The overestimation arising by scGW also agrees with the previous GW study.\cite{Shishkin2007}

Note that the important effect of core electrons on the valence-core interaction, and hence exchange self-energy has been discussed for bulk Si in the previous study.\cite{Ku2002} We have also evaluated the exchange self-energy elements of band edge states $\Gamma_v$ and $X_c$ with and without the core electrons. The difference in self-energy can be as large as 2 eV, in line with values given in that study.

We have also compared the spectral functions ($A_{j{\bf k}}(\omega)$) of the band edge states $\Gamma_v$ and $X_c$ of bulk Si from CPE to those obtained from Pade approximation, as shown in Figure~\ref{aw_plot} (a) and (b) for the cases of G$_0$W$_0$ and scGW, respectively. In the case of Si, results from these two approaches of analytic continuation are very similar in terms of peak position, as well as the broadening of peaks that is related to the lifetime of the associated quasiparticle states. 

\begin{table}[hpt]
\begin{center}
\begin{threeparttable}
\caption{Band gap values of bulk Si for various levels of approximation. The values in parentheses are computed using the Pade approximation.  All values are in eV.}
\begin{tabular}{ccc}
\hline
\hline
  \ \ & \ \ $\Gamma_v-\Gamma_c$ \ \ & \ \ $\Gamma_v-X_c$ \ \  \\
 \hline
 \ \ This work \ \ & & \\
 LDA & 2.52 & 0.58 \\
 G$_0$W$_0$ & \ \ 3.40 (3.38) \ \ & \ \ 1.38 (1.36) \ \ \\
 GW$_0$ & \ \ 3.69 (3.68) \ \ & \ \ 1.59 (1.58) \ \ \\
 scGW & \ \ 3.41 (3.57) \ \  & \ \ 1.44 (1.44) \ \ \\
\hline
plane-wave PP, G$_0$W$_0$\tnote{a} & 3.24 & 1.18 \\
\hline
all-electron, G$_0$W$_0$ & & \\
Hamada \textit{et al.}\tnote{b} & 3.30 & 1.14 \\
Kotani \textit{et al.}\tnote{c} & 3.13 & \\
Ku \textit{et al.}\tnote{d} & 3.12 & \\
Gomez-Abal \textit{et al.}\tnote{e} & & 1.15 \\
\hline
all-electron, scGW\tnote{d} & 3.48 & \\
\hline
\ \ Experiment\tnote{f} \ \ & 3.35 & 1.25 \\
\hline
\hline
\end{tabular}
\begin{tablenotes}
\item[a] {Reference~\onlinecite{Tiago2004}.}
\item[b] {Reference~\onlinecite{Hamada1990}.}
\item[c] {Reference~\onlinecite{Kotani2002}.}
\item[d] {Reference~\onlinecite{Ku2002}.}
\item[e] {Reference~\onlinecite{Gomez2008}.}
\item[f]{Reference~\onlinecite{Hellwege1987}.}
\end{tablenotes}
\label{Sigaps}
\end{threeparttable}
\end{center}
\end{table}

Finally, we demonstrate the computational advantage of the current implementation by evaluating silicon's G$_0$W$_0$ band gaps with a similar parameter set but using a direct numerical integration method in the real-frequency domain. We find that more than 1000 frequency points are needed to achieve the converged results. Since the computational load at each frequency/Matsubara-time grid point is similar, it is clear that significant computational speedup can be accomplished when GW calculation is performed in Matsubara-time domain (81 $\tau$ points used in this work).

\begin{table}[hpt]
\begin{center}
\begin{threeparttable}
\caption{Band gap values of bulk Ge for various levels of approximation. The values in parentheses are computed using the Pade approximation.  All values are in eV.}
\begin{tabular}{cccc}
\hline
\hline
  \ \ & \ \ $\Gamma_v-\Gamma_c$ \ \ & \ \ $\Gamma_v-L_c$  \ \ & \ \ $\Gamma_v-X_c$ \ \ \ \ \\
 \hline
 This work \ \ & & & \\
 LDA & -0.19 & 0.03 & 0.64 \\
 G$_0$W$_0$ & 0.49 (0.51) & 0.58 (0.59) & 0.65 (0.70) \\
 GW$_0$ & \ \ 1.09 (1.10) \ \ & \ \ 0.85 (0.86) \ \ & \ \ 1.35 (1.33) \ \ \\
 scGW & 1.11 (1.11) & 0.85 (0.85) & 1.30 (1.30) \\
\hline
plane-wave PP, G$_0$W$_0$\tnote{a} & 0.85 & 0.65 & 0.98 \\
\hline
all-electron, G$_0$W$_0$ & & & \\
Kotani \textit{et al.}\tnote{b} & 0.89 & 0.57 & \\
Ku \textit{et al.}\tnote{c} & 1.11 & 0.51 & 0.49 \\
\hline
all-electron, scGW\tnote{c} & 1.51 & 0.79 & 0.71 \\
\hline
\ \ Experiment\tnote{d} \ \ & 0.90 & 0.74 & 1.30 \\
\hline
\hline
\end{tabular}
\begin{tablenotes}
\item[a] {Reference~\onlinecite{Tiago2004}.}
\item[b] {Reference~\onlinecite{Kotani2002}.}
\item[c] {Reference~\onlinecite{Ku2002}.}
\item[d] {Reference~\onlinecite{Hellwege1987}.}
\end{tablenotes}
\label{Gegaps}
\end{threeparttable}
\end{center}
\end{table}

Table~\ref{Gegaps} summarizes the band gaps at different levels of theory for bulk Ge. The minimal, indirect band gap of bulk Ge is between $\Gamma_v$ and $L_c$ according to experiment.\cite{Hellwege1987} It is clear that both LDA and G$_0$W$_0$ predict a minimal band gap as direct at $\Gamma$, inconsistent with experiment. It is only when the self-consistency is considered in GW, (either GW$_0$ or scGW) that the correct indirect band gap can be predicted. Note that the results from scGW agree well with experimental data, and also very close to those from GW$_0$, regardless of CPE or Pade approximation being adopted. It is worth pointing out that there is a substantial difference between our results and those by Ku \textit{et al.}, with a band gap difference as large as 0.5 eV. We believe that such discrepancy is due mainly to the insufficient amount of empty bands used in their study, as pointed out in the previous study by Tiago \textit{et al.}.\cite{Tiago2004}

\begin{table}[hpt]
\begin{center}
\begin{threeparttable}
\caption{Band gap values of bulk GaAs for various levels of approximation. The values in parentheses are computed using the Pade approximation.  All values are in eV.}
\begin{tabular}{cccc}
\hline
\hline
  \ \ & \ \ $\Gamma_v-\Gamma_c$ \ \ & \ \ $\Gamma_v-L_c$  \ \ & \ \ $\Gamma_v-X_c$ \ \ \ \ \\
 \hline
 This work \ \ & & & \\
 LDA & 0.23 & 0.81 & 1.31 \\
 G$_0$W$_0$ & \ \ 1.48 (1.47) \ \ & \ \ 1.62 (1.62) \ \ & \ \ 1.98 (1.94) \ \ \\
 GW$_0$ & 1.82 (1.83) & 2.00 (2.00) & 2.31 (2.30) \\
 scGW & 1.80 (1.81) & 1.95 (1.96) & 2.23 (2.25) \\
\hline
plane-wave PP, G$_0$W$_0$\tnote{a} & 1.38 & 1.65 & 1.83 \\
\hline
all-electron, G$_0$W$_0$ & & & \\
Kotani \textit{et al.}\tnote{b} & 1.20 & 1.40 & 1.46 \\
Gomez-Abal \textit{et al.}\tnote{c} & 1.29 & & \\ 
Friedrich \textit{et al.}\tnote{d} & & & \\
\hline
\ \ Experiment\tnote{e} \ \ & 1.52 & 1.82 & 1.98 \\
\hline
\hline
\end{tabular}
\begin{tablenotes}
\item[a] {Reference~\onlinecite{Tiago2004}.}
\item[b] {Reference~\onlinecite{Kotani2002}.}
\item[c] {Reference~\onlinecite{Gomez2008}.}
\item[d] {Reference~\onlinecite{Friedrich2010}.}
\item[e] {Reference~\onlinecite{Aspnes1976}.}
\end{tablenotes}
\label{GaAsgaps}
\end{threeparttable}
\end{center}
\end{table}

Gallium aresnide is another common compound we use as a benchmark, with computed band gap results shown in Table~\ref{GaAsgaps}. This compound has also been extensively investigated, which has a direct electronic band gap at $\Gamma$. Our calculations show that G$_0$W$_0$ results in the best agreement with experiment,\cite{Aspnes1976} and also agree with previous all-electron G$_0$W$_0$ studies with a $\sim$0.2 eV difference. Moreover, both scGW and GW$_0$ lead to larger band gap values compared to the G$_0$W$_0$ results, and are overestimated by around 0.3 eV compared to experiment. Such trends regarding G$_0$W$_0$ and scGW are also in line with previous GW studies within the plane-wave PAW potential framework.\cite{Shishkin2007} Similar to the aforementioned compounds investigated, the CPE and Pade approximation lead to very close results to each other. Our scGW results presented here also serve as important predictions for this level of theory since there are no previous all-electron-based, self-consistent GW results for GaAs.

In general, G$_0$W$_0$ accurately predicts Si and GaAs band gap values but predicts inaccurate bulk Ge band gap values compared to experiment. On the other hand, scGW band gaps agree fairly well with experiment across all three elements, and GW$_0$ generally worsens the band gaps compared to scGW.

\subsection{Band gap calculations for other semiconductors and insulators}
\begin{figure*}[hpt]
\includegraphics[width=15cm]{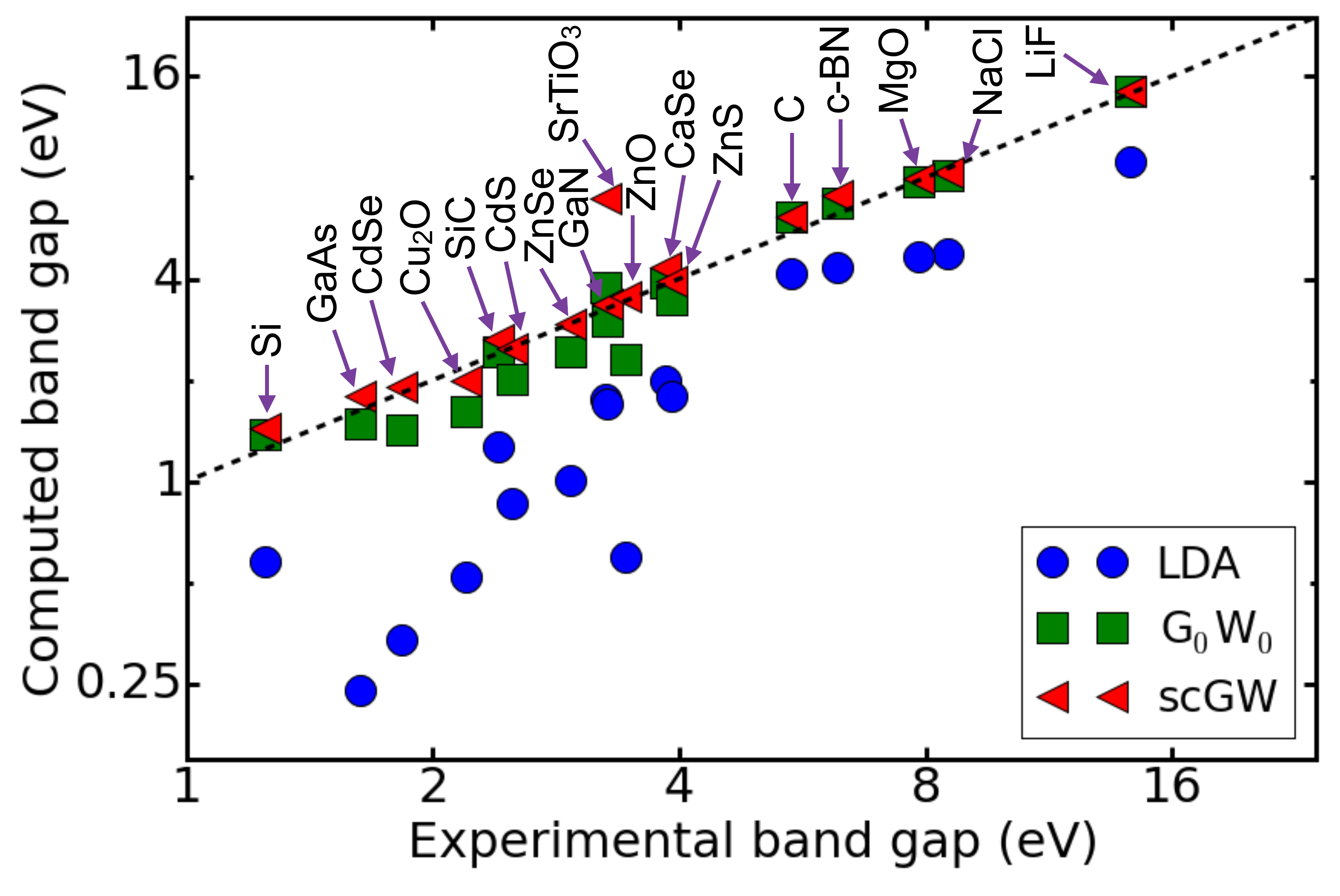}
\caption{(Color online) Computed electronic band gap at DFT-LDA as well as GW levels versus the experimental counterpart for all the compounds studied in this work except for Ge. Log scale is adopted for both axes.
}\label{gapsfig}
\end{figure*}

Having demonstrated the accuracy of scGW calculations for predicting electronic band gaps in benchmark materials, we next report results for 18 semiconductors/insulators that have band gaps covering a wide range of values from less than 1 eV to over 10 eV. The calculated minimal band gaps are summarized in Table~\ref{egs}, comparing all levels of approximation, and also in Fig.~\ref{gapsfig} which visualizes LDA, G$_0$W$_0$, and scGW results. As expected, the LDA band gaps are always severely underestimated compared to experimental values. Upon GW corrections, the electronic band gaps for all the systems studied are substantially improved. In the following, we discuss the effects of G$_0$W$_0$ and scGW band gap corrections by categorizing the compounds studied into three groups: (1) simple $s$-$p$ electron systems involving Si, SiC, C, BN, LiF, NaCl and MgO; (2) non-transition-metal systems with 3-$d$ electrons that include Ge, GaAs, GaN, CaSe, CdSe and CdS; and (3) transition-metal chalcogenides of \ce{SrTiO3}, \ce{Cu2O}, \ce{ZnO}, \ce{ZnS}, \ce{ZnSe}. 

\begin{table}[hpt]
\begin{center}
\begin{threeparttable}
\caption{Electronic band gap (in eV) of various semiconductors and insulators calculated by DFT-LDA, different levels of Matsubara-time GW (G$_0$W$_0$, GW$_0$ and scGW), and PPA-G$_0$W$_0$. Values in the parentheses are obtained using the Pade approximation. The experimental values (Expt.) are also given for comparison. }
\begin{tabular}{cccccccc}
\hline
\hline
  \ \ & \ \ LDA \ \ & \ \ G$_0$W$_0$  \ \ & \ \ GW$_0$ \ \ & \ \ full-GW \ \ & \ \ PPA-G$_0$W$_0$ \ \ & \ \ Expt. \ \ & \ \  \\
 \hline
 Si \ \ & 0.58 & 1.38 (1.36) & 1.59 (1.58) & 1.44 (1.44) & 1.28 & 1.25\tnote{a} &  \\
 Ge \ \ & 0.03 & 0.58 (0.59) & 0.85 (0.86) & 0.85 (0.85)  & 0.71 &  0.74\tnote{a} &  \\
 GaAs \ \ & 0.24 & 1.48 (1.47) & 1.82 (1.83) & 1.80 (1.81) & 1.51 & 1.52\tnote{b} &  \\
 SiC \ \ & 1.27 & 2.44 (2.45) & 2.90 (2.90) & 2.64 (2.56) & 2.30 & 2.40\tnote{a} &  \\
 CaSe \ \ & 2.00 & 3.89 (3.94) & 4.60 (4.64) & 4.35 (4.34) & 3.89 & 3.85\tnote{c} &  \\
 C \ \ & 4.14 & 6.15 (6.15) & 6.42 (6.43) & 6.10 (6.11) & 6.09 & 5.48\tnote{a} &  \\
 NaCl \ \ & 4.74 & 8.09 (8.11) & 9.00 (9.02) & 8.27 (8.28) & 8.11 & 8.5\tnote{d} &  \\
 MgO \ \ & 4.65 & 7.79 (7.78) & 8.74 (8.74) & 7.94 (7.94) & 7.75 & 7.83\tnote{e} &  \\
 BN \ \ & 4.34 & 6.71 (6.73) & 7.16 (7.18) & 7.10 (7.11) & 6.58 & 6.1-6.4\tnote{f} &  \\
 LiF \ \ & 8.94 & \ \ 14.51 (14.54) \ \ & \ \ 15.78 (15.81) \ \ & \ \ 14.45 (14.47) \ \ & 14.55 & 14.20\tnote{g} &  \\
 \ce{SrTiO3} \ \ & 1.75 & 3.58 (4.08) & 7.01 (7.13) & 6.87 (7.22) & 3.86 & 3.25\tnote{h} &  \\
 \ce{Cu2O} \ \ & 0.52 & 1.61 (1.54) & 2.16 (2.17) & 2.00 (2.02) & 1.59 & 2.17\tnote{i} &  \\
 GaN \ \ & 1.70 & 3.01 (3.05) & 3.61 (3.66) & 3.36 (3.38) & 3.03 & 3.27\tnote{j} &  \\
 ZnO \ \ & 0.60 & 2.31 (2.35) & 3.69 (3.71) & 3.53 (3.56) & 2.32 & 3.44\tnote{k} &  \\
 ZnS \ \ & 1.80 & 3.46 (3.43) & 4.06 (4.09) & 3.92 (3.85) & 3.43 & 3.91\tnote{k} &  \\
 ZnSe \ \ & 1.01 & 2.43 (2.48) & 3.03 (3.09) & 2.94 (2.96) & 2.50 & 2.95\tnote{a} &  \\
 CdSe \ \ & 0.34 & 1.42 (1.51) & 1.97 (1.98) & 1.92 (1.93) & 1.46 & 1.83\tnote{a} &  \\
 CdS \ \ & 0.86 & 2.01 (2.03) & 2.63 (2.66) & 2.49 (2.50) & 2.06 & 2.50\tnote{a} &  \\
\hline
\hline
\end{tabular}
\begin{tablenotes}
\item[a] {Reference~\onlinecite{Hellwege1987}.} 
\item[b] {Reference~\onlinecite{Aspnes1976}.}
\item[c] {Reference~\onlinecite{Kaneko1988}.}
\item[d] {Reference~\onlinecite{Poole1975}.}
\item[e] {Reference~\onlinecite{Whited1973}.}
\item[f] {Reference~\onlinecite{Levinshtein2001}.}
\item[g] {Reference~\onlinecite{Piacentini1976}.}
\item[h] {Reference~\onlinecite{vanBenthem2001}.}
\item[i] {Reference~\onlinecite{Baumeister1961}.}
\item[j] {Reference~\onlinecite{Okumura1994}.}
\item[k] {Reference~\onlinecite{Kittel1986}.}
\end{tablenotes}
\label{egs}
\end{threeparttable}
\end{center}
\end{table}

Concerning simple $s$-$p$ electron systems, the G$_0$W$_0$ corrected band gaps are in very good agreement with experimental data, with a relative band gap error of $\pm 10\%$ for most compounds with the exception of diamond, for which G$_0$W$_0$ overestimates the experimental gap by 0.6 eV ($12 \%$). This may be attributed to the RPA that leads to more severe underestimation of the screening effect in diamond, as pointed out in a previous study.\cite{Shishkin2007} Results using PPA are remarkably close to general G$_0$W$_0$ calculations, differing by only 0.1 eV or less. Our G$_0$W$_0$ band gaps are all comparable to previous all-electron G$_0$W$_0$ calculations.\cite{Friedrich2010,Gomez2008} When full self-consistency is taken into account, our calculations show that scGW may further overestimate the electronic band gap due probably to the underestimated screening effect by RPA, in agreement with previous findings.\cite{Shishkin2007,Shishkin2007a} The exceptions are diamond and the ionic crystals NaCl and LiF, for which the inclusion of self-consistency tends to improve results. It is worth pointing out that the band gaps from partially self-consistent GW$_0$ are considerably higher than scGW ones, which is in contrast to previous finding.\cite{Shishkin2007} This different trend is likely due to differences in method implementation. Specifically, the Green's function in our approach is fully updated during the GW$_0$ iteration, whereas in the previous GW$_0$ study only the shift in quasiparticle energies is updated in the Green's function.\cite{Shishkin2007} Compared to our scGW results, the further overestimation of GW$_0$ band gaps can be attributed to the further underestimation of screening due to the screened potential ($W$), which is not updated iteratively within GW$_0$. This also highlights the importance of the full self-consistency. Furthermore, the band gaps at different levels of GW have also been computed based on the CPE and Pade approximation. According to our results, they are in remarkable agreement with each other, with a typical difference of 0.1 eV or less in all the cases. This also confirms the applicability of the Pade approximation and the analytic continuation approach for $s$-$p$ electron systems. 

For non-transition-metal systems with 3-$d$ electrons, we have observed that G$_0$W$_0$ corrected band gaps are typically still underestimated. Fully self-consistent GW calculations are necessary to achieve better agreement with experimental data. The exception involves CaSe and GaAs, in which scGW leads to overestimated band gaps of about 0.3-0.5 eV, corresponding to a relative difference of more than 12 \%. Compared to scGW, GW$_0$ results in about 5 \% larger band gap values in the systems studied. The difference is smaller than in the case of $s$-$p$ electron systems. 

Moreover, in the cases of transition metal chalcogenides ZnO, ZnS and \ce{Cu2O}, it is clear that band gaps resulting from G$_0$W$_0$ are substantially underestimated by at least 0.5 eV compared to the experimental values. In particular, the G$_0$W$_0$ band gap of ZnO is 1 eV lower than the experimental data, agreeing well with previously underestimated values.\cite{Shishkin2007} Upon scGW, the band gaps of these systems are significantly improved such that they are within 0.2 eV of experimental results. In the case of the perovskite \ce{SrTiO3}, on the other hand, our G$_0$W$_0$ approach overestimates the band gap by about 0.3 eV, which is also consistent with the studies by Friedrich \textit{et al.}\cite{Friedrich2010} and Kang \textit{et al.},\cite{Kang2015} and both scGW and GW$_0$ worsen the band gap prediction further with a result of more than 6.8 eV, much higher than the experimental value of 3.25 eV. We have further varied parameters such as the number of conduction bands and the cutoff of reciprocal lattice vectors, and the corresponding results only change slightly.  

A previous GW study by Cappellini \textit{et al.} also showed that the minimal band gap of \ce{SrTiO3} can be severely overestimated even at the level of G$_0$W$_0$ (5.07 eV),\cite{Cappellini2000} Such an overestimation may be attributed to the improper description of local field effects by their model dielectric function. Moreover, our scGW band gap of \ce{SrTiO3} is indeed in line with the previous findings, in which the band gap is overestimated by around 0.9 eV in all-electron quasiparticle self-consistent GW,\cite{Schilfgaarde2006} whereas such overestimation becomes 1.8 eV in self-consistent GW with the diagonal approximation in the plane-wave PAW potential framework.\cite{Kang2015} Such a severe overestimation of the calculated scGW band gap is thus likely due to the poor accuracy of the diagonal approximation adopted for $G$, which leads to unchanged charge and spin densities during scGW. For systems with strongly correlated 3-$d$ electrons near the band edge, such as \ce{SrTiO3}, the quasiparticle wave functions may substantially deviate from K-S wave functions, resulting in considerable change in charge density and errors to the electronic band gap.  Future work will include an investigation into how the diagonal approximation affects electronic structure predictions of transition metal oxides and other strongly correlated systems.

Another possibility is the missing electron-hole correlation effects in RPA.\cite{Schilfgaarde2006} Such effects have proven to be crucial in conjunction with self-consistency to predict correct electronic band gaps.\cite{Shishkin2007a} Further investigation excluding the diagonal approximation and/or including screening effects beyond RPA is necessary and will be conducted in the future. Similar to the other two types of systems, the CPE and Pade approximation lead to similar band gaps differing within 0.05 eV. The only exception is \ce{SrTiO3},  for which the band gap from both approaches can differ by as much as 0.5 eV, as indicated in Table~\ref{egs} and shown via the spectral functions in Figure~\ref{aw_plot} (c) and (d). Regarding the spectral functions of band edge states, difference in weight of spectral functions indicates that the estimated lifetime of the quasiparticle states may differ substantially. CPE appears to be the more valid method for analytic continuation given its general agreement with experiment for a wide range of systems.  Still, the applicability of Pade approximation is justified for many systems based on our calculations.

\section{Conclusion\label{conclusion}}
To summarize, we have implemented an efficient Matsubara-time GW approach in conjunction with CPE, a newly developed analytic continuation method. The method has been used in a detailed study of the electronic band gaps across 18 semiconductors and/or insulators at the levels of G$_0$W$_0$, GW$_0$ and scGW approximations. Benchmark calculations of silicon's electronic structure demonstrate the accuracy and computational speedup of our Matsubara-time method compared to previously used, frequency-domain calculations.  Our results demonstrate that for most of the simple $s$-$p$ electron systems, G$_0$W$_0$ leads to reasonable agreement with experiments, and scGW tends to overestimate the calculated band gaps, whereas scGW is required for more accurate band gaps in the cases of 3-$d$ transition metal chalcogenides. These findings are in line with the previous GW studies and it is likely due to the underestimated screening effects by RPA during scGW. We have also found that the band gap of strongly correlated systems such as \ce{SrTiO3} can be substantially overestimated within the current framework, and off-diagonal elements in $G$ as well as the electron-hole correlation effects beyond RPA may need to be included for more accurate results in those systems. Moreover, we have compared the results from both CPE and Pade approximation. In general, CPE results are more consistently in agreement with experimental data in a wide range of systems, suggesting the applicability of CPE for analytic continuation as a standard for GW calculations.

\section{Acknowledgements}
This work is supported by the U.S. Department of Energy (DOE), Office of Basic Energy Sciences (BES), under Contract No. DE-FG02-02ER45995. A. G. Eguiluz acknowledges funding support from National Science Foundation (NSF) under Grant No. OCI-0904972. All calculations were performed at the National Energy Research Scientific Computing Center (NERSC).

\bibliography{scGW_paper}

\begin{thebibliography}{65}
\expandafter\ifx\csname natexlab\endcsname\relax\def\natexlab#1{#1}\fi
\expandafter\ifx\csname bibnamefont\endcsname\relax
  \def\bibnamefont#1{#1}\fi
\expandafter\ifx\csname bibfnamefont\endcsname\relax
  \def\bibfnamefont#1{#1}\fi
\expandafter\ifx\csname citenamefont\endcsname\relax
  \def\citenamefont#1{#1}\fi
\expandafter\ifx\csname url\endcsname\relax
  \def\url#1{\texttt{#1}}\fi
\expandafter\ifx\csname urlprefix\endcsname\relax\def\urlprefix{URL }\fi
\providecommand{\bibinfo}[2]{#2}
\providecommand{\eprint}[2][]{\url{#2}}

\bibitem[{\citenamefont{Kohn and Sham}(1965)}]{Kohn1965}
\bibinfo{author}{\bibfnamefont{W.}~\bibnamefont{Kohn}} \bibnamefont{and}
  \bibinfo{author}{\bibfnamefont{L.~J.} \bibnamefont{Sham}},
  \bibinfo{journal}{Phys. Rev.} \textbf{\bibinfo{volume}{140}},
  \bibinfo{pages}{A1133} (\bibinfo{year}{1965}).

\bibitem[{\citenamefont{Hohenberg and Kohn}(1964)}]{Hohenberg1964}
\bibinfo{author}{\bibfnamefont{P.}~\bibnamefont{Hohenberg}} \bibnamefont{and}
  \bibinfo{author}{\bibfnamefont{W.}~\bibnamefont{Kohn}},
  \bibinfo{journal}{Phys. Rev.} \textbf{\bibinfo{volume}{136}},
  \bibinfo{pages}{B864} (\bibinfo{year}{1964}).

\bibitem[{\citenamefont{Kohn et~al.}(1996)\citenamefont{Kohn, Becke, and
  Parr}}]{Kohn1996}
\bibinfo{author}{\bibfnamefont{W.}~\bibnamefont{Kohn}},
  \bibinfo{author}{\bibfnamefont{A.~D.} \bibnamefont{Becke}}, \bibnamefont{and}
  \bibinfo{author}{\bibfnamefont{R.~G.} \bibnamefont{Parr}},
  \bibinfo{journal}{J. Phys. Chem.} \textbf{\bibinfo{volume}{100}},
  \bibinfo{pages}{12974} (\bibinfo{year}{1996}).

\bibitem[{\citenamefont{Dreizler and Gross}(2012)}]{Dreizler2012}
\bibinfo{author}{\bibfnamefont{R.~M.} \bibnamefont{Dreizler}} \bibnamefont{and}
  \bibinfo{author}{\bibfnamefont{E.~K.} \bibnamefont{Gross}},
  \emph{\bibinfo{title}{Density functional theory: an approach to the quantum
  many-body problem}} (\bibinfo{publisher}{Springer Science \& Business Media},
  \bibinfo{year}{2012}).

\bibitem[{\citenamefont{Norskov et~al.}(2011)\citenamefont{Norskov,
  Abild-Pedersen, Studt, and Bligaard}}]{Norskov2011}
\bibinfo{author}{\bibfnamefont{J.~K.} \bibnamefont{Norskov}},
  \bibinfo{author}{\bibfnamefont{F.}~\bibnamefont{Abild-Pedersen}},
  \bibinfo{author}{\bibfnamefont{F.}~\bibnamefont{Studt}}, \bibnamefont{and}
  \bibinfo{author}{\bibfnamefont{T.}~\bibnamefont{Bligaard}},
  \bibinfo{journal}{Proc. Natl. Acad. Sci.} \textbf{\bibinfo{volume}{108}},
  \bibinfo{pages}{937} (\bibinfo{year}{2011}).

\bibitem[{\citenamefont{Ruiz et~al.}(2012)\citenamefont{Ruiz, Liu, Zojer,
  Scheffler, and Tkatchenko}}]{Ruiz2012}
\bibinfo{author}{\bibfnamefont{V.~G.} \bibnamefont{Ruiz}},
  \bibinfo{author}{\bibfnamefont{W.}~\bibnamefont{Liu}},
  \bibinfo{author}{\bibfnamefont{E.}~\bibnamefont{Zojer}},
  \bibinfo{author}{\bibfnamefont{M.}~\bibnamefont{Scheffler}},
  \bibnamefont{and}
  \bibinfo{author}{\bibfnamefont{A.}~\bibnamefont{Tkatchenko}},
  \bibinfo{journal}{Phys. Rev. Lett.} \textbf{\bibinfo{volume}{108}},
  \bibinfo{pages}{146103} (\bibinfo{year}{2012}).

\bibitem[{\citenamefont{Sham and Schl\"uter}(1983)}]{Sham1983}
\bibinfo{author}{\bibfnamefont{L.~J.} \bibnamefont{Sham}} \bibnamefont{and}
  \bibinfo{author}{\bibfnamefont{M.}~\bibnamefont{Schl\"uter}},
  \bibinfo{journal}{Phys. Rev. Lett.} \textbf{\bibinfo{volume}{51}},
  \bibinfo{pages}{1888} (\bibinfo{year}{1983}).

\bibitem[{\citenamefont{Hedin}(1965)}]{Hedin1965}
\bibinfo{author}{\bibfnamefont{L.}~\bibnamefont{Hedin}},
  \bibinfo{journal}{Phys. Rev.} \textbf{\bibinfo{volume}{139}},
  \bibinfo{pages}{A796} (\bibinfo{year}{1965}).

\bibitem[{\citenamefont{Eshuis et~al.}(2012)\citenamefont{Eshuis, Bates, and
  Furche}}]{Eshuis2012}
\bibinfo{author}{\bibfnamefont{H.}~\bibnamefont{Eshuis}},
  \bibinfo{author}{\bibfnamefont{J.}~\bibnamefont{Bates}}, \bibnamefont{and}
  \bibinfo{author}{\bibfnamefont{F.}~\bibnamefont{Furche}},
  \bibinfo{journal}{Theor. Chem. Acc.} \textbf{\bibinfo{volume}{131}},
  \bibinfo{eid}{1084} (\bibinfo{year}{2012}).

\bibitem[{\citenamefont{Ren et~al.}(2012)\citenamefont{Ren, Rinke, Joas, and
  Scheffler}}]{Ren2012}
\bibinfo{author}{\bibfnamefont{X.}~\bibnamefont{Ren}},
  \bibinfo{author}{\bibfnamefont{P.}~\bibnamefont{Rinke}},
  \bibinfo{author}{\bibfnamefont{C.}~\bibnamefont{Joas}}, \bibnamefont{and}
  \bibinfo{author}{\bibfnamefont{M.}~\bibnamefont{Scheffler}},
  \bibinfo{journal}{J. Mater. Sci.} \textbf{\bibinfo{volume}{47}},
  \bibinfo{pages}{7447} (\bibinfo{year}{2012}).

\bibitem[{\citenamefont{Chu et~al.}(2014)\citenamefont{Chu, Kozhevnikov,
  Schulthess, and Cheng}}]{Chu2014}
\bibinfo{author}{\bibfnamefont{I.-H.} \bibnamefont{Chu}},
  \bibinfo{author}{\bibfnamefont{A.}~\bibnamefont{Kozhevnikov}},
  \bibinfo{author}{\bibfnamefont{T.~C.} \bibnamefont{Schulthess}},
  \bibnamefont{and} \bibinfo{author}{\bibfnamefont{H.-P.} \bibnamefont{Cheng}},
  \bibinfo{journal}{J. Chem. Phys.} \textbf{\bibinfo{volume}{141}},
  \bibinfo{eid}{044709} (\bibinfo{year}{2014}).

\bibitem[{\citenamefont{Yang et~al.}(2007)\citenamefont{Yang, Park, Son, Cohen,
  and Louie}}]{Yang2007}
\bibinfo{author}{\bibfnamefont{L.}~\bibnamefont{Yang}},
  \bibinfo{author}{\bibfnamefont{C.-H.} \bibnamefont{Park}},
  \bibinfo{author}{\bibfnamefont{Y.-W.} \bibnamefont{Son}},
  \bibinfo{author}{\bibfnamefont{M.~L.} \bibnamefont{Cohen}}, \bibnamefont{and}
  \bibinfo{author}{\bibfnamefont{S.~G.} \bibnamefont{Louie}},
  \bibinfo{journal}{Phys. Rev. Lett.} \textbf{\bibinfo{volume}{99}},
  \bibinfo{pages}{186801} (\bibinfo{year}{2007}).

\bibitem[{\citenamefont{Shishkin and Kresse}(2007)}]{Shishkin2007}
\bibinfo{author}{\bibfnamefont{M.}~\bibnamefont{Shishkin}} \bibnamefont{and}
  \bibinfo{author}{\bibfnamefont{G.}~\bibnamefont{Kresse}},
  \bibinfo{journal}{Phys. Rev. B} \textbf{\bibinfo{volume}{75}},
  \bibinfo{pages}{235102} (\bibinfo{year}{2007}).

\bibitem[{\citenamefont{Kresse et~al.}(2012)\citenamefont{Kresse, Marsman,
  Hintzsche, and Flage-Larsen}}]{Kresse2012}
\bibinfo{author}{\bibfnamefont{G.}~\bibnamefont{Kresse}},
  \bibinfo{author}{\bibfnamefont{M.}~\bibnamefont{Marsman}},
  \bibinfo{author}{\bibfnamefont{L.~E.} \bibnamefont{Hintzsche}},
  \bibnamefont{and}
  \bibinfo{author}{\bibfnamefont{E.}~\bibnamefont{Flage-Larsen}},
  \bibinfo{journal}{Phys. Rev. B} \textbf{\bibinfo{volume}{85}},
  \bibinfo{pages}{045205} (\bibinfo{year}{2012}).

\bibitem[{\citenamefont{Fuchs et~al.}(2007)\citenamefont{Fuchs, Furthm\"uller,
  Bechstedt, Shishkin, and Kresse}}]{Fuchs2007}
\bibinfo{author}{\bibfnamefont{F.}~\bibnamefont{Fuchs}},
  \bibinfo{author}{\bibfnamefont{J.}~\bibnamefont{Furthm\"uller}},
  \bibinfo{author}{\bibfnamefont{F.}~\bibnamefont{Bechstedt}},
  \bibinfo{author}{\bibfnamefont{M.}~\bibnamefont{Shishkin}}, \bibnamefont{and}
  \bibinfo{author}{\bibfnamefont{G.}~\bibnamefont{Kresse}},
  \bibinfo{journal}{Phys. Rev. B} \textbf{\bibinfo{volume}{76}},
  \bibinfo{pages}{115109} (\bibinfo{year}{2007}).

\bibitem[{\citenamefont{Faleev et~al.}(2004)\citenamefont{Faleev, van
  Schilfgaarde, and Kotani}}]{Faleev2004}
\bibinfo{author}{\bibfnamefont{S.~V.} \bibnamefont{Faleev}},
  \bibinfo{author}{\bibfnamefont{M.}~\bibnamefont{van Schilfgaarde}},
  \bibnamefont{and} \bibinfo{author}{\bibfnamefont{T.}~\bibnamefont{Kotani}},
  \bibinfo{journal}{Phys. Rev. Lett.} \textbf{\bibinfo{volume}{93}},
  \bibinfo{pages}{126406} (\bibinfo{year}{2004}).

\bibitem[{\citenamefont{Deslippe et~al.}(2012)\citenamefont{Deslippe,
  Samsonidze, Strubbe, Jain, Cohen, and Louie}}]{Deslippe2012}
\bibinfo{author}{\bibfnamefont{J.}~\bibnamefont{Deslippe}},
  \bibinfo{author}{\bibfnamefont{G.}~\bibnamefont{Samsonidze}},
  \bibinfo{author}{\bibfnamefont{D.~A.} \bibnamefont{Strubbe}},
  \bibinfo{author}{\bibfnamefont{M.}~\bibnamefont{Jain}},
  \bibinfo{author}{\bibfnamefont{M.~L.} \bibnamefont{Cohen}}, \bibnamefont{and}
  \bibinfo{author}{\bibfnamefont{S.~G.} \bibnamefont{Louie}},
  \bibinfo{journal}{Comput. Phys. Commun.} \textbf{\bibinfo{volume}{183}},
  \bibinfo{pages}{1269 } (\bibinfo{year}{2012}).

\bibitem[{\citenamefont{Marini et~al.}(2001)\citenamefont{Marini, Onida, and
  Del~Sole}}]{Marini2001}
\bibinfo{author}{\bibfnamefont{A.}~\bibnamefont{Marini}},
  \bibinfo{author}{\bibfnamefont{G.}~\bibnamefont{Onida}}, \bibnamefont{and}
  \bibinfo{author}{\bibfnamefont{R.}~\bibnamefont{Del~Sole}},
  \bibinfo{journal}{Phys. Rev. Lett.} \textbf{\bibinfo{volume}{88}},
  \bibinfo{pages}{016403} (\bibinfo{year}{2001}).

\bibitem[{\citenamefont{Dixit et~al.}(2011)\citenamefont{Dixit, Saniz, Lamoen,
  and Partoens}}]{Dixit2011}
\bibinfo{author}{\bibfnamefont{H.}~\bibnamefont{Dixit}},
  \bibinfo{author}{\bibfnamefont{R.}~\bibnamefont{Saniz}},
  \bibinfo{author}{\bibfnamefont{D.}~\bibnamefont{Lamoen}}, \bibnamefont{and}
  \bibinfo{author}{\bibfnamefont{B.}~\bibnamefont{Partoens}},
  \bibinfo{journal}{Comput. Phys. Commun.} \textbf{\bibinfo{volume}{182}},
  \bibinfo{pages}{2029 } (\bibinfo{year}{2011}).

\bibitem[{\citenamefont{Pham et~al.}(2013)\citenamefont{Pham, Nguyen, Rocca,
  and Galli}}]{Pham2013}
\bibinfo{author}{\bibfnamefont{T.~A.} \bibnamefont{Pham}},
  \bibinfo{author}{\bibfnamefont{H.-V.} \bibnamefont{Nguyen}},
  \bibinfo{author}{\bibfnamefont{D.}~\bibnamefont{Rocca}}, \bibnamefont{and}
  \bibinfo{author}{\bibfnamefont{G.}~\bibnamefont{Galli}},
  \bibinfo{journal}{Phys. Rev. B} \textbf{\bibinfo{volume}{87}},
  \bibinfo{pages}{155148} (\bibinfo{year}{2013}).

\bibitem[{\citenamefont{Ku and Eguiluz}(2002)}]{Ku2002}
\bibinfo{author}{\bibfnamefont{W.}~\bibnamefont{Ku}} \bibnamefont{and}
  \bibinfo{author}{\bibfnamefont{A.~G.} \bibnamefont{Eguiluz}},
  \bibinfo{journal}{Phys. Rev. Lett.} \textbf{\bibinfo{volume}{89}},
  \bibinfo{pages}{126401} (\bibinfo{year}{2002}).

\bibitem[{\citenamefont{Sharma et~al.}(2005)\citenamefont{Sharma, Dewhurst, and
  Ambrosch-Draxl}}]{Sharma2005}
\bibinfo{author}{\bibfnamefont{S.}~\bibnamefont{Sharma}},
  \bibinfo{author}{\bibfnamefont{J.~K.} \bibnamefont{Dewhurst}},
  \bibnamefont{and}
  \bibinfo{author}{\bibfnamefont{C.}~\bibnamefont{Ambrosch-Draxl}},
  \bibinfo{journal}{Phys. Rev. Lett.} \textbf{\bibinfo{volume}{95}},
  \bibinfo{pages}{136402} (\bibinfo{year}{2005}).

\bibitem[{\citenamefont{Friedrich et~al.}(2010)\citenamefont{Friedrich,
  Bl\"ugel, and Schindlmayr}}]{Friedrich2010}
\bibinfo{author}{\bibfnamefont{C.}~\bibnamefont{Friedrich}},
  \bibinfo{author}{\bibfnamefont{S.}~\bibnamefont{Bl\"ugel}}, \bibnamefont{and}
  \bibinfo{author}{\bibfnamefont{A.}~\bibnamefont{Schindlmayr}},
  \bibinfo{journal}{Phys. Rev. B} \textbf{\bibinfo{volume}{81}},
  \bibinfo{pages}{125102} (\bibinfo{year}{2010}).

\bibitem[{\citenamefont{Jiang et~al.}(2013)\citenamefont{Jiang, Gómez-Abal,
  Li, Meisenbichler, Ambrosch-Draxl, and Scheffler}}]{Jiang2013}
\bibinfo{author}{\bibfnamefont{H.}~\bibnamefont{Jiang}},
  \bibinfo{author}{\bibfnamefont{R.~I.} \bibnamefont{Gómez-Abal}},
  \bibinfo{author}{\bibfnamefont{X.-Z.} \bibnamefont{Li}},
  \bibinfo{author}{\bibfnamefont{C.}~\bibnamefont{Meisenbichler}},
  \bibinfo{author}{\bibfnamefont{C.}~\bibnamefont{Ambrosch-Draxl}},
  \bibnamefont{and}
  \bibinfo{author}{\bibfnamefont{M.}~\bibnamefont{Scheffler}},
  \bibinfo{journal}{Comput. Phys. Commun.} \textbf{\bibinfo{volume}{184}},
  \bibinfo{pages}{348 } (\bibinfo{year}{2013}).

\bibitem[{\citenamefont{Usuda et~al.}(2002)\citenamefont{Usuda, Hamada, Kotani,
  and van Schilfgaarde}}]{Usuda2002}
\bibinfo{author}{\bibfnamefont{M.}~\bibnamefont{Usuda}},
  \bibinfo{author}{\bibfnamefont{N.}~\bibnamefont{Hamada}},
  \bibinfo{author}{\bibfnamefont{T.}~\bibnamefont{Kotani}}, \bibnamefont{and}
  \bibinfo{author}{\bibfnamefont{M.}~\bibnamefont{van Schilfgaarde}},
  \bibinfo{journal}{Phys. Rev. B} \textbf{\bibinfo{volume}{66}},
  \bibinfo{pages}{125101} (\bibinfo{year}{2002}).

\bibitem[{\citenamefont{Bl\"ochl}(1994)}]{Bochl1994}
\bibinfo{author}{\bibfnamefont{P.~E.} \bibnamefont{Bl\"ochl}},
  \bibinfo{journal}{Phys. Rev. B} \textbf{\bibinfo{volume}{50}},
  \bibinfo{pages}{17953} (\bibinfo{year}{1994}).

\bibitem[{\citenamefont{Shishkin and Kresse}(2006)}]{Shishkin2006}
\bibinfo{author}{\bibfnamefont{M.}~\bibnamefont{Shishkin}} \bibnamefont{and}
  \bibinfo{author}{\bibfnamefont{G.}~\bibnamefont{Kresse}},
  \bibinfo{journal}{Phys. Rev. B} \textbf{\bibinfo{volume}{74}},
  \bibinfo{pages}{035101} (\bibinfo{year}{2006}).

\bibitem[{\citenamefont{G\'omez-Abal et~al.}(2008)\citenamefont{G\'omez-Abal,
  Li, Scheffler, and Ambrosch-Draxl}}]{Gomez2008}
\bibinfo{author}{\bibfnamefont{R.}~\bibnamefont{G\'omez-Abal}},
  \bibinfo{author}{\bibfnamefont{X.}~\bibnamefont{Li}},
  \bibinfo{author}{\bibfnamefont{M.}~\bibnamefont{Scheffler}},
  \bibnamefont{and}
  \bibinfo{author}{\bibfnamefont{C.}~\bibnamefont{Ambrosch-Draxl}},
  \bibinfo{journal}{Phys. Rev. Lett.} \textbf{\bibinfo{volume}{101}},
  \bibinfo{pages}{106404} (\bibinfo{year}{2008}).

\bibitem[{\citenamefont{Li et~al.}(2012)\citenamefont{Li, Gómez-Abal, Jiang,
  Ambrosch-Draxl, and Scheffler}}]{Li2012}
\bibinfo{author}{\bibfnamefont{X.-Z.} \bibnamefont{Li}},
  \bibinfo{author}{\bibfnamefont{R.}~\bibnamefont{Gómez-Abal}},
  \bibinfo{author}{\bibfnamefont{H.}~\bibnamefont{Jiang}},
  \bibinfo{author}{\bibfnamefont{C.}~\bibnamefont{Ambrosch-Draxl}},
  \bibnamefont{and}
  \bibinfo{author}{\bibfnamefont{M.}~\bibnamefont{Scheffler}},
  \bibinfo{journal}{New J. Phys.} \textbf{\bibinfo{volume}{14}},
  \bibinfo{pages}{023006} (\bibinfo{year}{2012}).

\bibitem[{\citenamefont{Baym and Kadanoff}(1961)}]{Baym1961}
\bibinfo{author}{\bibfnamefont{G.}~\bibnamefont{Baym}} \bibnamefont{and}
  \bibinfo{author}{\bibfnamefont{L.~P.} \bibnamefont{Kadanoff}},
  \bibinfo{journal}{Phys. Rev.} \textbf{\bibinfo{volume}{124}},
  \bibinfo{pages}{287} (\bibinfo{year}{1961}).

\bibitem[{\citenamefont{Baym}(1962)}]{Baym1962}
\bibinfo{author}{\bibfnamefont{G.}~\bibnamefont{Baym}}, \bibinfo{journal}{Phys.
  Rev.} \textbf{\bibinfo{volume}{127}}, \bibinfo{pages}{1391}
  (\bibinfo{year}{1962}).

\bibitem[{\citenamefont{Dahlen et~al.}(2006)\citenamefont{Dahlen, van Leeuwen,
  and von Barth}}]{Dahlen2006}
\bibinfo{author}{\bibfnamefont{N.~E.} \bibnamefont{Dahlen}},
  \bibinfo{author}{\bibfnamefont{R.}~\bibnamefont{van Leeuwen}},
  \bibnamefont{and} \bibinfo{author}{\bibfnamefont{U.}~\bibnamefont{von
  Barth}}, \bibinfo{journal}{Phys. Rev. A} \textbf{\bibinfo{volume}{73}},
  \bibinfo{pages}{012511} (\bibinfo{year}{2006}).

\bibitem[{\citenamefont{Rinke et~al.}(2005)\citenamefont{Rinke, Qteish,
  Neugebauer, Freysoldt, and Scheffler}}]{Rinke2005}
\bibinfo{author}{\bibfnamefont{P.}~\bibnamefont{Rinke}},
  \bibinfo{author}{\bibfnamefont{A.}~\bibnamefont{Qteish}},
  \bibinfo{author}{\bibfnamefont{J.}~\bibnamefont{Neugebauer}},
  \bibinfo{author}{\bibfnamefont{C.}~\bibnamefont{Freysoldt}},
  \bibnamefont{and}
  \bibinfo{author}{\bibfnamefont{M.}~\bibnamefont{Scheffler}},
  \bibinfo{journal}{New J. Phys.} \textbf{\bibinfo{volume}{7}},
  \bibinfo{pages}{126} (\bibinfo{year}{2005}).

\bibitem[{\citenamefont{Caruso et~al.}(2012)\citenamefont{Caruso, Rinke, Ren,
  Scheffler, and Rubio}}]{Carus2012}
\bibinfo{author}{\bibfnamefont{F.}~\bibnamefont{Caruso}},
  \bibinfo{author}{\bibfnamefont{P.}~\bibnamefont{Rinke}},
  \bibinfo{author}{\bibfnamefont{X.}~\bibnamefont{Ren}},
  \bibinfo{author}{\bibfnamefont{M.}~\bibnamefont{Scheffler}},
  \bibnamefont{and} \bibinfo{author}{\bibfnamefont{A.}~\bibnamefont{Rubio}},
  \bibinfo{journal}{Phys. Rev. B} \textbf{\bibinfo{volume}{86}},
  \bibinfo{pages}{081102} (\bibinfo{year}{2012}).

\bibitem[{\citenamefont{Caruso et~al.}(2013)\citenamefont{Caruso, Rinke, Ren,
  Rubio, and Scheffler}}]{Caruso2013}
\bibinfo{author}{\bibfnamefont{F.}~\bibnamefont{Caruso}},
  \bibinfo{author}{\bibfnamefont{P.}~\bibnamefont{Rinke}},
  \bibinfo{author}{\bibfnamefont{X.}~\bibnamefont{Ren}},
  \bibinfo{author}{\bibfnamefont{A.}~\bibnamefont{Rubio}}, \bibnamefont{and}
  \bibinfo{author}{\bibfnamefont{M.}~\bibnamefont{Scheffler}},
  \bibinfo{journal}{Phys. Rev. B} \textbf{\bibinfo{volume}{88}},
  \bibinfo{pages}{075105} (\bibinfo{year}{2013}).

\bibitem[{\citenamefont{Mahan}(2013)}]{Mahan2013}
\bibinfo{author}{\bibfnamefont{G.~D.} \bibnamefont{Mahan}},
  \emph{\bibinfo{title}{Many-particle physics}} (\bibinfo{publisher}{Springer
  Science \& Business Media}, \bibinfo{year}{2013}).

\bibitem[{\citenamefont{Bruus and Flensberg}(2004)}]{Bruus2004}
\bibinfo{author}{\bibfnamefont{H.}~\bibnamefont{Bruus}} \bibnamefont{and}
  \bibinfo{author}{\bibfnamefont{K.}~\bibnamefont{Flensberg}},
  \emph{\bibinfo{title}{Many-body quantum theory in condensed matter physics:
  an introduction}} (\bibinfo{publisher}{OUP Oxford}, \bibinfo{year}{2004}).

\bibitem[{\citenamefont{Kutepov et~al.}(2009)\citenamefont{Kutepov, Savrasov,
  and Kotliar}}]{Kutepov2009}
\bibinfo{author}{\bibfnamefont{A.}~\bibnamefont{Kutepov}},
  \bibinfo{author}{\bibfnamefont{S.~Y.} \bibnamefont{Savrasov}},
  \bibnamefont{and} \bibinfo{author}{\bibfnamefont{G.}~\bibnamefont{Kotliar}},
  \bibinfo{journal}{Phys. Rev. B} \textbf{\bibinfo{volume}{80}},
  \bibinfo{pages}{041103} (\bibinfo{year}{2009}).

\bibitem[{\citenamefont{Vidberg and Serene}(1977)}]{Vidberg1977}
\bibinfo{author}{\bibfnamefont{H.}~\bibnamefont{Vidberg}} \bibnamefont{and}
  \bibinfo{author}{\bibfnamefont{J.}~\bibnamefont{Serene}},
  \bibinfo{journal}{J. Low Temp. Phys.} \textbf{\bibinfo{volume}{29}},
  \bibinfo{pages}{179} (\bibinfo{year}{1977}).

\bibitem[{\citenamefont{Staar et~al.}(2014)\citenamefont{Staar, Ydens,
  Kozhevnikov, Locquet, and Schulthess}}]{Staar2014}
\bibinfo{author}{\bibfnamefont{P.}~\bibnamefont{Staar}},
  \bibinfo{author}{\bibfnamefont{B.}~\bibnamefont{Ydens}},
  \bibinfo{author}{\bibfnamefont{A.}~\bibnamefont{Kozhevnikov}},
  \bibinfo{author}{\bibfnamefont{J.-P.} \bibnamefont{Locquet}},
  \bibnamefont{and}
  \bibinfo{author}{\bibfnamefont{T.}~\bibnamefont{Schulthess}},
  \bibinfo{journal}{Phys. Rev. B} \textbf{\bibinfo{volume}{89}},
  \bibinfo{pages}{245114} (\bibinfo{year}{2014}).

\bibitem[{\citenamefont{Kozhevnikov et~al.}(2010)\citenamefont{Kozhevnikov,
  Eguiluz, and Schulthess}}]{Kozhevnikov2010}
\bibinfo{author}{\bibfnamefont{A.}~\bibnamefont{Kozhevnikov}},
  \bibinfo{author}{\bibfnamefont{A.~G.} \bibnamefont{Eguiluz}},
  \bibnamefont{and} \bibinfo{author}{\bibfnamefont{T.~C.}
  \bibnamefont{Schulthess}}, in \emph{\bibinfo{booktitle}{SC'10 Proceedings of
  the 2010 ACM/IEEE International Conference for High Performance Computing,
  Networking, Storage, and Analysis (IEEE Computer Society, Washington, DC,
  2010)}} (\bibinfo{year}{2010}), pp. \bibinfo{pages}{1--10}.

\bibitem[{exc()}]{excitingplus}
\bibinfo{howpublished}{See \url{https://code.google.com/p/exciting-plus/} for
  information about the source code.}

\bibitem[{\citenamefont{Stan et~al.}(2009)\citenamefont{Stan, Dahlen, and van
  Leeuwen}}]{Stan2009}
\bibinfo{author}{\bibfnamefont{A.}~\bibnamefont{Stan}},
  \bibinfo{author}{\bibfnamefont{N.~E.} \bibnamefont{Dahlen}},
  \bibnamefont{and} \bibinfo{author}{\bibfnamefont{R.}~\bibnamefont{van
  Leeuwen}}, \bibinfo{journal}{J. Chem. Phys.} \textbf{\bibinfo{volume}{130}},
  \bibinfo{eid}{114105} (\bibinfo{year}{2009}).

\bibitem[{\citenamefont{Sj{\"o}stedt et~al.}(2000)\citenamefont{Sj{\"o}stedt,
  Nordstr{\"o}m, and Singh}}]{Sjostedt2000}
\bibinfo{author}{\bibfnamefont{E.}~\bibnamefont{Sj{\"o}stedt}},
  \bibinfo{author}{\bibfnamefont{L.}~\bibnamefont{Nordstr{\"o}m}},
  \bibnamefont{and} \bibinfo{author}{\bibfnamefont{D.}~\bibnamefont{Singh}},
  \bibinfo{journal}{Solid state commun.} \textbf{\bibinfo{volume}{114}},
  \bibinfo{pages}{15} (\bibinfo{year}{2000}).

\bibitem[{\citenamefont{Perdew and Wang}(1992)}]{Perdew1992}
\bibinfo{author}{\bibfnamefont{J.~P.} \bibnamefont{Perdew}} \bibnamefont{and}
  \bibinfo{author}{\bibfnamefont{Y.}~\bibnamefont{Wang}},
  \bibinfo{journal}{Phys. Rev. B} \textbf{\bibinfo{volume}{45}},
  \bibinfo{pages}{13244} (\bibinfo{year}{1992}).

\bibitem[{\citenamefont{Godby and Needs}(1989)}]{Godby1989}
\bibinfo{author}{\bibfnamefont{R.~W.} \bibnamefont{Godby}} \bibnamefont{and}
  \bibinfo{author}{\bibfnamefont{R.~J.} \bibnamefont{Needs}},
  \bibinfo{journal}{Phys. Rev. Lett.} \textbf{\bibinfo{volume}{62}},
  \bibinfo{pages}{1169} (\bibinfo{year}{1989}).

\bibitem[{\citenamefont{Stankovski et~al.}(2011)\citenamefont{Stankovski,
  Antonius, Waroquiers, Miglio, Dixit, Sankaran, Giantomassi, Gonze, C\^ot\'e,
  and Rignanese}}]{Stankovski2011}
\bibinfo{author}{\bibfnamefont{M.}~\bibnamefont{Stankovski}},
  \bibinfo{author}{\bibfnamefont{G.}~\bibnamefont{Antonius}},
  \bibinfo{author}{\bibfnamefont{D.}~\bibnamefont{Waroquiers}},
  \bibinfo{author}{\bibfnamefont{A.}~\bibnamefont{Miglio}},
  \bibinfo{author}{\bibfnamefont{H.}~\bibnamefont{Dixit}},
  \bibinfo{author}{\bibfnamefont{K.}~\bibnamefont{Sankaran}},
  \bibinfo{author}{\bibfnamefont{M.}~\bibnamefont{Giantomassi}},
  \bibinfo{author}{\bibfnamefont{X.}~\bibnamefont{Gonze}},
  \bibinfo{author}{\bibfnamefont{M.}~\bibnamefont{C\^ot\'e}}, \bibnamefont{and}
  \bibinfo{author}{\bibfnamefont{G.-M.} \bibnamefont{Rignanese}},
  \bibinfo{journal}{Phys. Rev. B} \textbf{\bibinfo{volume}{84}},
  \bibinfo{pages}{241201} (\bibinfo{year}{2011}).

\bibitem[{\citenamefont{Landolt-B{\"o}rnstein}(1987)}]{Hellwege1987}
\bibinfo{author}{\bibnamefont{Landolt-B{\"o}rnstein}},
  \emph{\bibinfo{title}{Numerical Data and Functional Relationships in Science
  and Technology, edited by K.-H. Hellwege, O. Madelung, M. Schulz, and H.
  Weiss, New Series}}, vol. \bibinfo{volume}{III}
  (\bibinfo{publisher}{Springer-Verlag, New York}, \bibinfo{year}{1987}).

\bibitem[{\citenamefont{Tiago et~al.}(2004)\citenamefont{Tiago, Ismail-Beigi,
  and Louie}}]{Tiago2004}
\bibinfo{author}{\bibfnamefont{M.~L.} \bibnamefont{Tiago}},
  \bibinfo{author}{\bibfnamefont{S.}~\bibnamefont{Ismail-Beigi}},
  \bibnamefont{and} \bibinfo{author}{\bibfnamefont{S.~G.} \bibnamefont{Louie}},
  \bibinfo{journal}{Phys. Rev. B} \textbf{\bibinfo{volume}{69}},
  \bibinfo{pages}{125212} (\bibinfo{year}{2004}).

\bibitem[{\citenamefont{Hamada et~al.}(1990)\citenamefont{Hamada, Hwang, and
  Freeman}}]{Hamada1990}
\bibinfo{author}{\bibfnamefont{N.}~\bibnamefont{Hamada}},
  \bibinfo{author}{\bibfnamefont{M.}~\bibnamefont{Hwang}}, \bibnamefont{and}
  \bibinfo{author}{\bibfnamefont{A.~J.} \bibnamefont{Freeman}},
  \bibinfo{journal}{Phys. Rev. B} \textbf{\bibinfo{volume}{41}},
  \bibinfo{pages}{3620} (\bibinfo{year}{1990}).

\bibitem[{\citenamefont{Kotani and van Schilfgaarde}(2002)}]{Kotani2002}
\bibinfo{author}{\bibfnamefont{T.}~\bibnamefont{Kotani}} \bibnamefont{and}
  \bibinfo{author}{\bibfnamefont{M.}~\bibnamefont{van Schilfgaarde}},
  \bibinfo{journal}{Solid State Commun.} \textbf{\bibinfo{volume}{121}},
  \bibinfo{pages}{461 } (\bibinfo{year}{2002}).

\bibitem[{\citenamefont{Aspnes}(1976)}]{Aspnes1976}
\bibinfo{author}{\bibfnamefont{D.~E.} \bibnamefont{Aspnes}},
  \bibinfo{journal}{Phys. Rev. B} \textbf{\bibinfo{volume}{14}},
  \bibinfo{pages}{5331} (\bibinfo{year}{1976}).

\bibitem[{\citenamefont{Kaneko and Koda}(1988)}]{Kaneko1988}
\bibinfo{author}{\bibfnamefont{Y.}~\bibnamefont{Kaneko}} \bibnamefont{and}
  \bibinfo{author}{\bibfnamefont{T.}~\bibnamefont{Koda}}, \bibinfo{journal}{J.
  Cryst. Growth} \textbf{\bibinfo{volume}{86}}, \bibinfo{pages}{72}
  (\bibinfo{year}{1988}).

\bibitem[{\citenamefont{Poole et~al.}(1975)\citenamefont{Poole, Jenkin,
  Liesegang, and Leckey}}]{Poole1975}
\bibinfo{author}{\bibfnamefont{R.~T.} \bibnamefont{Poole}},
  \bibinfo{author}{\bibfnamefont{J.~G.} \bibnamefont{Jenkin}},
  \bibinfo{author}{\bibfnamefont{J.}~\bibnamefont{Liesegang}},
  \bibnamefont{and} \bibinfo{author}{\bibfnamefont{R.~C.~G.}
  \bibnamefont{Leckey}}, \bibinfo{journal}{Phys. Rev. B}
  \textbf{\bibinfo{volume}{11}}, \bibinfo{pages}{5179} (\bibinfo{year}{1975}).

\bibitem[{\citenamefont{Whited et~al.}(1973)\citenamefont{Whited, Flaten, and
  Walker}}]{Whited1973}
\bibinfo{author}{\bibfnamefont{R.}~\bibnamefont{Whited}},
  \bibinfo{author}{\bibfnamefont{C.~J.} \bibnamefont{Flaten}},
  \bibnamefont{and} \bibinfo{author}{\bibfnamefont{W.}~\bibnamefont{Walker}},
  \bibinfo{journal}{Solid State Commun.} \textbf{\bibinfo{volume}{13}},
  \bibinfo{pages}{1903 } (\bibinfo{year}{1973}).

\bibitem[{Lev(2001)}]{Levinshtein2001}
\emph{\bibinfo{title}{Properties of Advanced Semiconductor Materials: GaN, AlN,
  InN, BN, SiC, and SiGe, edited by M. E. Levinshtein, S. L. Rumyantsev, and M.
  S. Shur}} (\bibinfo{publisher}{Wiley, New York}, \bibinfo{year}{2001}).

\bibitem[{\citenamefont{Piacentini et~al.}(1976)\citenamefont{Piacentini,
  Lynch, and Olson}}]{Piacentini1976}
\bibinfo{author}{\bibfnamefont{M.}~\bibnamefont{Piacentini}},
  \bibinfo{author}{\bibfnamefont{D.~W.} \bibnamefont{Lynch}}, \bibnamefont{and}
  \bibinfo{author}{\bibfnamefont{C.~G.} \bibnamefont{Olson}},
  \bibinfo{journal}{Phys. Rev. B} \textbf{\bibinfo{volume}{13}},
  \bibinfo{pages}{5530} (\bibinfo{year}{1976}).

\bibitem[{\citenamefont{van Benthem et~al.}(2001)\citenamefont{van Benthem,
  Elsässer, and French}}]{vanBenthem2001}
\bibinfo{author}{\bibfnamefont{K.}~\bibnamefont{van Benthem}},
  \bibinfo{author}{\bibfnamefont{C.}~\bibnamefont{Elsässer}},
  \bibnamefont{and} \bibinfo{author}{\bibfnamefont{R.~H.}
  \bibnamefont{French}}, \bibinfo{journal}{J. Appl. Phys.}
  \textbf{\bibinfo{volume}{90}}, \bibinfo{pages}{6156} (\bibinfo{year}{2001}).

\bibitem[{\citenamefont{Baumeister}(1961)}]{Baumeister1961}
\bibinfo{author}{\bibfnamefont{P.~W.} \bibnamefont{Baumeister}},
  \bibinfo{journal}{Phys. Rev.} \textbf{\bibinfo{volume}{121}},
  \bibinfo{pages}{359} (\bibinfo{year}{1961}).

\bibitem[{\citenamefont{Okumura et~al.}(1994)\citenamefont{Okumura, Yoshida,
  and Okahisa}}]{Okumura1994}
\bibinfo{author}{\bibfnamefont{H.}~\bibnamefont{Okumura}},
  \bibinfo{author}{\bibfnamefont{S.}~\bibnamefont{Yoshida}}, \bibnamefont{and}
  \bibinfo{author}{\bibfnamefont{T.}~\bibnamefont{Okahisa}},
  \bibinfo{journal}{Appl. Phys. Lett.} \textbf{\bibinfo{volume}{64}},
  \bibinfo{pages}{2997} (\bibinfo{year}{1994}).

\bibitem[{\citenamefont{Kittel and McEuen}(1986)}]{Kittel1986}
\bibinfo{author}{\bibfnamefont{C.}~\bibnamefont{Kittel}} \bibnamefont{and}
  \bibinfo{author}{\bibfnamefont{P.}~\bibnamefont{McEuen}},
  \emph{\bibinfo{title}{Introduction to solid state physics}},
  vol.~\bibinfo{volume}{8} (\bibinfo{publisher}{Wiley New York},
  \bibinfo{year}{1986}).

\bibitem[{\citenamefont{Shishkin et~al.}(2007)\citenamefont{Shishkin, Marsman,
  and Kresse}}]{Shishkin2007a}
\bibinfo{author}{\bibfnamefont{M.}~\bibnamefont{Shishkin}},
  \bibinfo{author}{\bibfnamefont{M.}~\bibnamefont{Marsman}}, \bibnamefont{and}
  \bibinfo{author}{\bibfnamefont{G.}~\bibnamefont{Kresse}},
  \bibinfo{journal}{Phys. Rev. Lett.} \textbf{\bibinfo{volume}{99}},
  \bibinfo{pages}{246403} (\bibinfo{year}{2007}).

\bibitem[{\citenamefont{Kang et~al.}(2015)\citenamefont{Kang, Kang, and
  Han}}]{Kang2015}
\bibinfo{author}{\bibfnamefont{G.}~\bibnamefont{Kang}},
  \bibinfo{author}{\bibfnamefont{Y.}~\bibnamefont{Kang}}, \bibnamefont{and}
  \bibinfo{author}{\bibfnamefont{S.}~\bibnamefont{Han}},
  \bibinfo{journal}{Phys. Rev. B} \textbf{\bibinfo{volume}{91}},
  \bibinfo{pages}{155141} (\bibinfo{year}{2015}).

\bibitem[{\citenamefont{Cappellini et~al.}(2000)\citenamefont{Cappellini,
  Bouette-Russo, Amadon, Noguera, and Finocchi}}]{Cappellini2000}
\bibinfo{author}{\bibfnamefont{G.}~\bibnamefont{Cappellini}},
  \bibinfo{author}{\bibfnamefont{S.}~\bibnamefont{Bouette-Russo}},
  \bibinfo{author}{\bibfnamefont{B.}~\bibnamefont{Amadon}},
  \bibinfo{author}{\bibfnamefont{C.}~\bibnamefont{Noguera}}, \bibnamefont{and}
  \bibinfo{author}{\bibfnamefont{F.}~\bibnamefont{Finocchi}},
  \bibinfo{journal}{J. Phys.: Condens. Matter} \textbf{\bibinfo{volume}{12}},
  \bibinfo{pages}{3671} (\bibinfo{year}{2000}).

\bibitem[{\citenamefont{van Schilfgaarde et~al.}(2006)\citenamefont{van
  Schilfgaarde, Kotani, and Faleev}}]{Schilfgaarde2006}
\bibinfo{author}{\bibfnamefont{M.}~\bibnamefont{van Schilfgaarde}},
  \bibinfo{author}{\bibfnamefont{T.}~\bibnamefont{Kotani}}, \bibnamefont{and}
  \bibinfo{author}{\bibfnamefont{S.}~\bibnamefont{Faleev}},
  \bibinfo{journal}{Phys. Rev. Lett.} \textbf{\bibinfo{volume}{96}},
  \bibinfo{pages}{226402} (\bibinfo{year}{2006}).

\end{thebibliography}

\end{document}